# MSci Project Report: Kinetic Monte Carlo simulations inspired by epitaxial graphene growth

Jonathan Lloyd-Williams

28th April 2011

Supervisor: Professor Dimitri Vvedensky

Assessor: Professor Matthew Foulkes

## Abstract

Graphene, a flat monolayer of carbon atoms packed tightly into a two dimensional hexagonal lattice, has unusual electronic properties which have many promising nanoelectronic applications. Epitaxial growth of graphene is the only viable route for producing graphene for electronic applications but the atomistic growth mechanisms are not well understood. Recent Low Energy Electron Microscopy (LEEM) experiments show that the growth velocity of 2D graphene islands on Ru(0001) varies with the fifth power of the supersaturation of carbon adatoms. This suggests that graphene islands grow by the addition of clusters of five atoms rather than by the usual mechanism of single adatom attachment. A simple rate theory based on this result supposes that graphene islands only form when six five-atom clusters collide. This theory is able to quantitatively account for all data measured in the LEEM experiments.

We have carried out Kinetic Monte Carlo (KMC) simulations in order to further investigate the general scenario of epitaxial growth by the attachment of mobile clusters of atoms. We did not seek to directly replicate the Gr/Ru(0001) system but instead considered a model involving mobile tetramers of atoms on a square lattice. Our results show that the energy barrier for tetramer break up and the number of tetramers that must collide in order to nucleate an immobile island are the important parameters for determining whether, as in the Gr/Ru(0001) system, the adatom density at the onset of island nucleation is an increasing function of temperature. A relatively large energy barrier for adatom attachment to islands is required in order for our model to produce an equilibrium adatom density that is a large fraction of the nucleation density. A large energy barrier for tetramer attachment to islands is also needed for the island density to dramatically decrease with increasing temperature. We show that islands grow with a velocity that varies with the fourth power of the supersaturation of adatoms when tetramer attachment is the dominant process for island growth.

# Contents

# 1. Graphene

Graphene is a flat monolayer of carbon atoms packed tightly into a two-dimensional (2D) hexagonal lattice.[1] Graphene has been studied theoretically for over sixty years since it is important in understanding the electronic properties of carbon allotropes such as graphite or carbon nanotubes[2] but until recently was not thought to exist in a free state because it was believed that thermal fluctuations would render the 2D sheets unstable. The discovery of free standing graphene was first reported in 2004 by Novoselov and Geim[3] who used micromechanical cleavage to isolate graphene from bulk graphite crystals. Novoselov and Geim were able to identify the graphene because of the optical effect it creates on top of a $SiO_2$ substrate making it visible in an optical microscope.[1] On 5th October 2010 the Nobel Prize in Physics was awarded to Andre Geim and Konstantin Novoselov for their discovery of isolated graphene.

The physical properties of graphene were recently reviewed in reference 2. The 2D layer is formed by $sp^2$-hybridized carbon atoms arranged in a hexagonal lattice and separated by 1.42Å. The hybrid orbitals give rise to σ bonds between atoms. The remaining orbital, $p_z$, is perpendicular to the planar structure and forms covalent bonds between neighbouring carbon atoms, leading to a half filled π band. This band can be described using a tight-binding approach as was first done by Wallace in 1947 in his work on the band theory of graphite.[4] The resulting band structure has the dispersion relation,

$$E_{\pm}(q) \approx \pm v_F q + O\left(\left(q / k_F\right)^2\right) \qquad (1)$$

about the Fermi points in reciprocal space so that q = |**k** – $\mathbf{k}_F$| and $v_F$ is the constant Fermi velocity ≈ $10^6$ $ms^{-1}$. The signs + and – refer to the upper and lower π bands respectively. Graphene is a zero gap semiconductor or semi metal so the Fermi surface consists of only six points and, because the above dispersion relation is of the same form as that of ultrarelativistic particles described by the massless Dirac equation, they are commonly called the Dirac points. This property determines most of the singular physical characteristics of free standing graphene and it means that graphene is a laboratory condensed matter system to test (2+1)-dimensional quantum electrodynamics.[5]

The experimental observation of some of these properties has confirmed the existence of charge carriers in graphene described by the massless Dirac equation.[6] For example, the carriers' cyclotron mass depends on the square root of the density of states and the integer quantum Hall effect occurs at half-integer filling factor, these both being characteristics of massless Dirac carriers.

The physical properties of graphene result in charge carriers that can be tuned continuously between electrons and holes to very high concentrations of up to $10^{13}$ $cm^{-2}$.[3] The carrier mobilities can exceed 15,000 $cm^2V^{-1}s^{-1}$ and have a weak temperature dependence giving the potential that they can be significantly increased by reducing impurity scattering.[1] The mobilities of carriers in graphene also remain high in electrically and chemically doped samples in contrast to conventional semiconductors.[1] This means that graphene has the potential to become a replacement for silicon in the electronics industry as the high mobility of graphene in doped devices facilitates the construction of so-called ballistic transistors at room temperature.[1] IBM researchers have recently reported the use of graphene to make transistors in the GHz scale with performances surpassing those of similar silicon transistors.[7,8]

Transistors and diodes require the presence of a band-gap for their operation so standard graphene sheets are not feasible for this application. However, research has been carried out into alternative graphene based structures with the presence of a band-gap. For example graphene nanoribbons[9,10] have a band gap proportional to the ribbon width and graphene nanomeshes,[11] which are made of graphene sheets punched with an array of nanoscale holes, also have a band gap.

Although graphene was first produced using micromechanical cleavage by Novoselov *et al.* they have conceded that epitaxial growth of graphene offers probably the only viable route towards electronic applications.[1] Epitaxial growth is the name given to the process of growing a thin film of material in a particular crystallographic orientation relationship to a substrate layer.[12] The adsorbed form of graphene on metal surfaces has been known for 40 years but unsurprisingly interest has recently been renewed in this area with a review published in 2009 summarising recent progress.[13]

## 2. Experimental evidence for graphene growth by carbon cluster attachment

The initial motivation for this work was provided by the results of a Low Energy Electron Microscopy (LEEM) study[14,15] carried out by Loginova and co-workers at Sandia National Laboratories in the USA who investigated the nucleation and growth of graphene on the transition metal ruthenium (Ru) by measuring the local concentration of carbon adatoms from which graphene forms and the growth rates of individual graphene islands.

The Sandia group used the Ru(0001) surface as a substrate because it supports the growth of large graphene films and because the ruthenium electronic band structure is such that low energy electrons reflect strongly from the Ru(0001) surface. Carbon atoms were deposited onto the surface both by heating a high purity carbon rod and by exposing the surface to ethylene ($C_2H_4$). During graphene growth the local concentration of mobile carbon adatoms was determined from the intensity of the LEEM images formed from the reflected electron beam. This allowed for the simultaneous imaging of graphene islands and measurement of the adatom concentration.

The essential measurement of the growth rate of graphene islands is the velocity at which the edges of the islands advance across the substrate, v, as a function of the supersaturation of carbon adatoms on the surface, $n - n_{eq}$, where n is the density of adatoms and $n_{eq}$ is the density in equilibrium with graphene. The average island edge velocity is given by,

$$v = \frac{1}{P}\frac{dA}{dt} \tag{2}$$

where P is the island perimeter and A is the island area. Figure 1 shows that the growth velocity of graphene islands measured by the Sandia group was found to be a highly nonlinear function of the supersaturation. During epitaxial growth of thin films the growth of islands is typically due to a small imbalance in the rate at which atoms attach and detach from the island edge and hence the edge velocity is linear in adatom supersaturation.[14] Loginova and co-workers suggested that the nonlinear growth rate found for graphene islands is due to the energy barrier for single carbon atom attachment to graphene islands being much larger than the energy required for a cluster of i carbon atoms to form and then to attach. Assuming that the growth rate of islands is then proportional to how far the density of these multi-atom clusters is out of equilibrium then the island edge velocity is of the form,

$$v = B\left[\left(\frac{n}{n_{eq}}\right)^{i} - 1\right] \tag{3}$$

The Sandia group fitted this equation to the experimental data varying three parameters: B, $n_{eq}$ and i. The values of $n_{eq}$ obtained from the fits were equal, within experimental errors, to the equilibrium adatom densities directly measured during the experiments which illustrates the self-consistency of the data and the growth model. Figure 1 shows that a growth velocity of the form given by equation 3 accurately describes a range of experimental data. For eight sets of data over temperatures ranging from 740K to 1070K the fits to equation 3 gave i = 4.8±0.5. An Arrhenius plot of the fitted values of B gives the activation energy for cluster attachment to graphene to be 2.0±0.1eV.

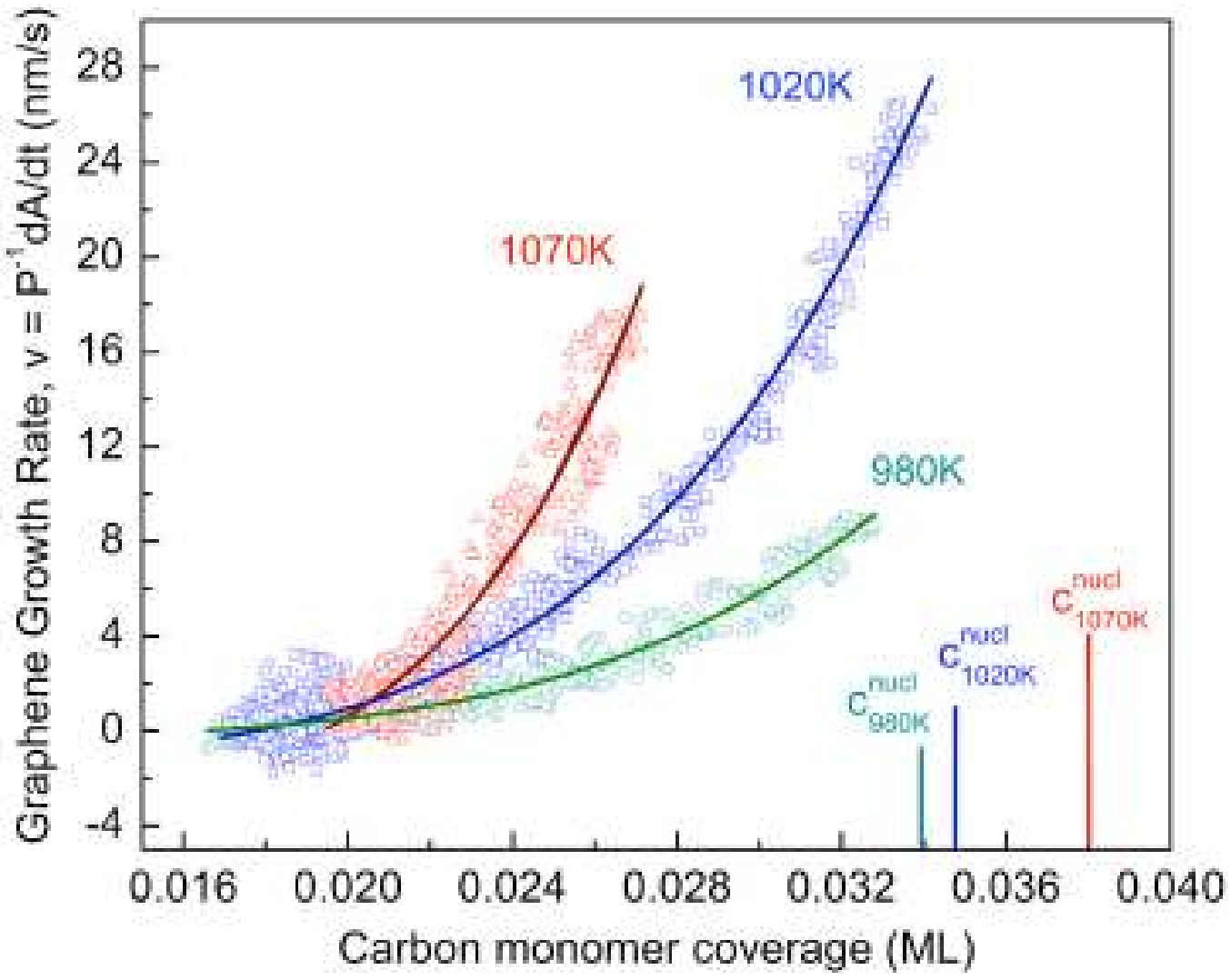

**Figure 1.** Island growth velocity as a function of adatom density. The solid lines are fits to equation 3. The vertical lines show the adatom density at island nucleation. From reference 14.

In seeking to understand why graphene appears to grow by clusters of about five atoms Loginova and co-workers carried out density functional theory (DFT) calculations and found that carbon adatoms on Ru (0001) lie 1.0Å above the plane through the centres of the surface Ru atoms. Graphene islands lie at heights of up to 3.7Å above the surface plane.[16] Carbon adatoms must break three bonds to the Ru substrate and simultaneously form several new bonds to the carbon atoms in the graphene islands in order to have a low energy barrier to attach to graphene. Loginova and co-workers suggested that the significant physical distance between the adatoms and the graphene islands may mean that this process is difficult and that a thermally excited intermediate state is required for attachment. This multi-atom cluster state involves both C-C and C-Ru bonds allowing the carbon atoms to bridge the spatial and energetic gap between the adatoms and the graphene. The Sandia group suggested that clusters may be formed uniformly over the Ru terraces or may only be formed during the attachment event itself.

Figure 2 shows that at least three carbon atoms must be added to a compact graphene island to form a new six-membered ring. However, adding three carbon atoms produces an isolated ring (shown in blue) which Loginova and co-workers argue may not provide a pathway to attachment since three of the carbon atoms in the new ring are only bonded to two other carbon atoms. Attaching five carbon atoms to the compact island adds two adjacent six-membered rings (shown in red) and only two of the carbon atoms in the new rings only have bonds to two other carbon atoms. Loginova and co-workers suggest that if this configuration is the smallest stable nucleus for further island growth then adding fewer than five carbon atoms would not grow graphene. Once this stable nucleus forms additional rings can potentially be formed from groups of two and three carbon atoms but it is the attachment of five-atom clusters that is the limiting step for graphene island growth.

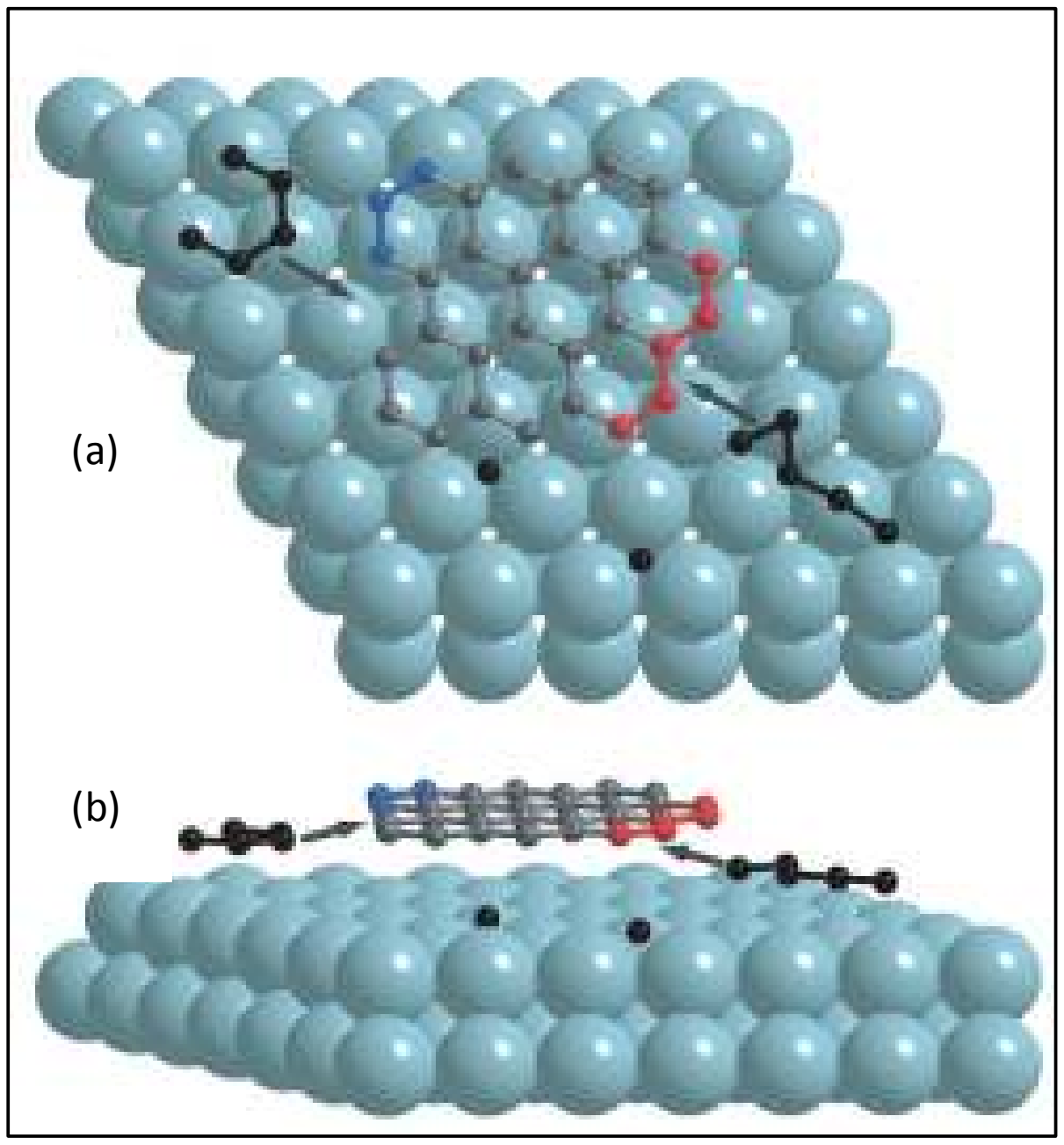

**Figure 2.** Schematic (a) top and (b) side views of graphene growth by attachment of carbon atom clusters. Two five atom clusters (black) are shown. At least three carbon atoms must be added to the graphene island (grey shaded atoms) to form a new six-membered ring. Adding three (blue) or five (red) carbon atoms forms one or two new six-membered rings, respectively. From reference 14.

Figure 3 shows the temperature dependence of the nucleation density of carbon adatoms compared with the equilibrium adatom density. We can see that the nucleation density, $n_{nuc}$, is approximately twice the equilibrium density, $n_{eq}$, across the temperature range shown. It is also interesting to note that the nucleation density is an increasing function of temperature. In typical epitaxial systems the limiting step for island nucleation is the rate of adatom diffusion and hence $n_{nuc}$ decreases with temperature as long as the critical island is approximately temperature independent. The Sandia group argue that the result $n_{nuc} \approx 2n_{eq}$ means that the critical island size is roughly constant for graphene formation.

Loginova and co-workers also observed that the density of nucleated graphene islands decreases rapidly with increasing temperature.[15] They attribute this feature to the large activation energy of cluster attachment to islands which means that at low temperatures the density of adatoms remains large after the initial nucleation of graphene islands and there is more time for further islands to nucleate. However there has not been any data published relating to this result. The Sandia group have also extended many of the conclusions presented here to the Gr/Ir(111) system.[15]

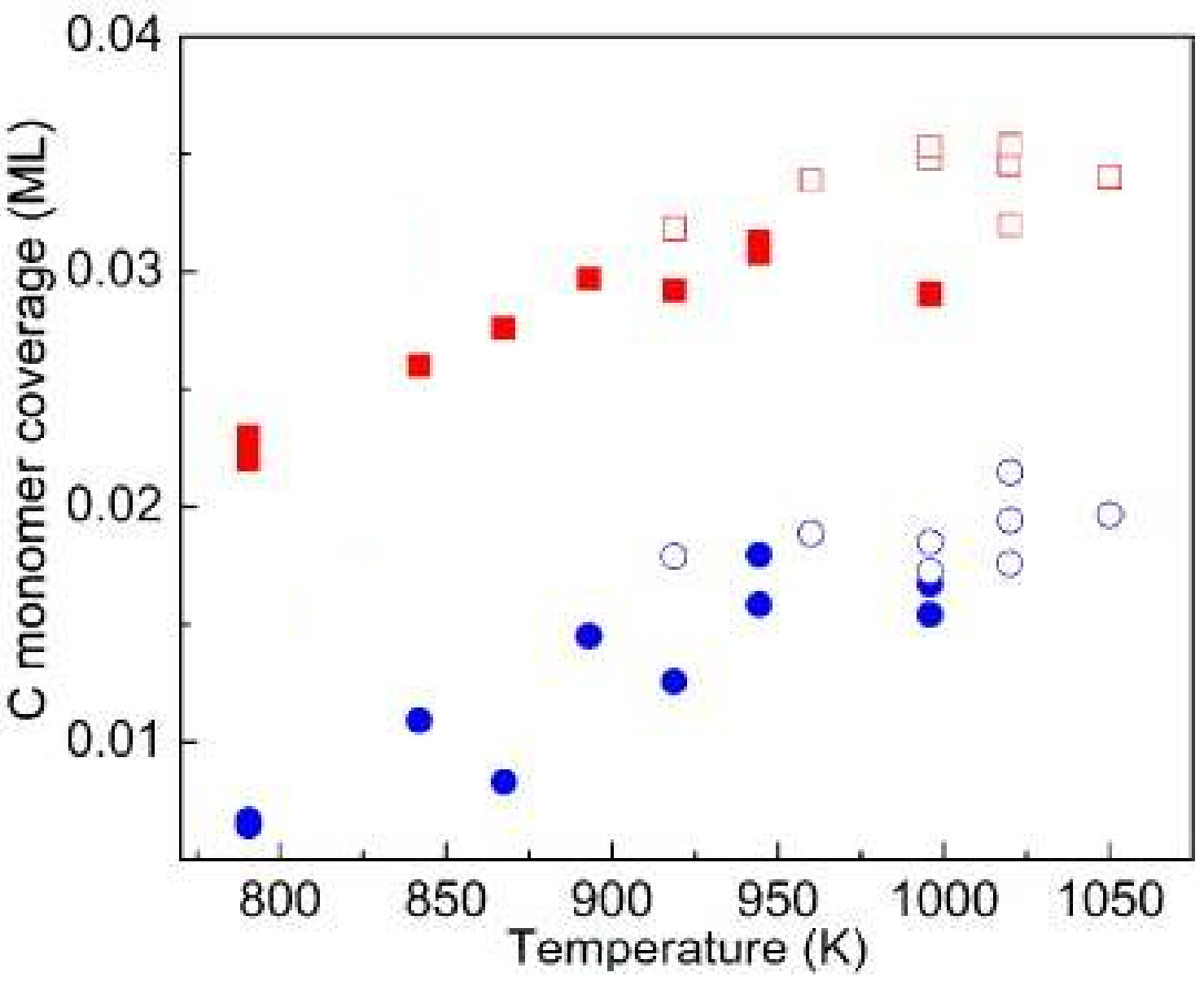

**Figure 3.** Temperature dependence of the carbon adatom concentration in equilibrium with graphene, $n_{eq}$ (circles), and carbon adatom concentration needed to nucleate graphene, $n_{nuc}$ (squares), on Ru(0001) after carbon-vapour (filled symbols) and ethylene (hollow symbols) deposition. From reference 15.

## 3. A rate theory of epitaxial graphene growth on metal surfaces

The work of Loginova and co-workers[14,15] motivated Zangwill and Vvedensky to develop a rate theory for the epitaxial growth of graphene on metal surfaces.[17] Rate equations have been used for some time for modelling epitaxial systems and provide the conceptual framework for interpreting essentially all epitaxial growth scenarios.[12,18]

In this model carbon atoms arrive at the surface with flux F and move across the surface with diffusion constant D. When i=5 carbon adatoms collide they form a mobile cluster which moves across the surface with diffusion constant D'. These mobile clusters have a finite lifetime and break up into five adatoms with rate K. When j=6 of the mobile clusters collide an immobile island of size i x j is formed. Immobile islands grow by capturing additional clusters as well as single adatoms. Single atoms detach from immobile islands with rate K'. The islands formed from six mobile clusters may not themselves immediately transform into graphene but instead pass through a series of intermediate structures.

The value i=5 was derived from experiment while the value of j=6 was obtained by Zangwill and Vvedensky from a fit to the experimental data of the temperature dependence of $n_{nuc}$.

The rate equations are expressed in terms of the homogeneous densities of carbon adatoms n, mobile five atom clusters c, and immobile islands N, as,

$$\frac{dn}{dt} = F - iDn^i + iKc - DnN + K'N \tag{4}$$

$$\frac{dc}{dt} = Dn^i - D'cN - jD'c^j - Kc \tag{5}$$

$$\frac{dN}{dt} = D'c^j \tag{6}$$

This set of equations does not include any spatial information in the form of capture numbers but Zangwill and Vvedensky found that the equations contain enough physical content to quantitatively account for the measured data contained in references 14 and 15. Each of the parameters D, D', K and K' in the rate equations have the Arrhenius form $\nu_0 e^{-\beta E}$, in which the attempt frequency $\nu_0 = 2k_BT/h$, $k_B$ is Boltzmann's constant, T is the absolute temperature, h is Planck's constant, $\beta = 1/k_BT$ and E is the energy barrier to the process.

There are five undetermined parameters in the rate equations: the four energy barriers and the value of j. However, by considering the constraints applied by temperature dependence of $n_{nuc}$ and $n_{eq}$ reported by the Sandia group, the optimization of the rate equations was reduced to a three parameter fit.

The time evolution of the densities of the carbon adatoms, the five-atom clusters and the immobile islands given by the solutions of the rate equations with the optimized parameters are shown in figure 4. The adatom and island density profiles are qualitatively different from the profiles given by typical epitaxial systems in which only adatoms are mobile and there is a critical size for island formation.[19] The presence of the mobile five-atom clusters which determine the critical island size clearly must be responsible for this behaviour. We can see that the five-atom clusters show a sudden increase in their density only when there is a sufficient adatom density for clusters to form. The nucleation of islands only occurs when the density of clusters is high enough for a six cluster collision to become likely. Once the island density is established it remains essentially constant with no further nucleation.

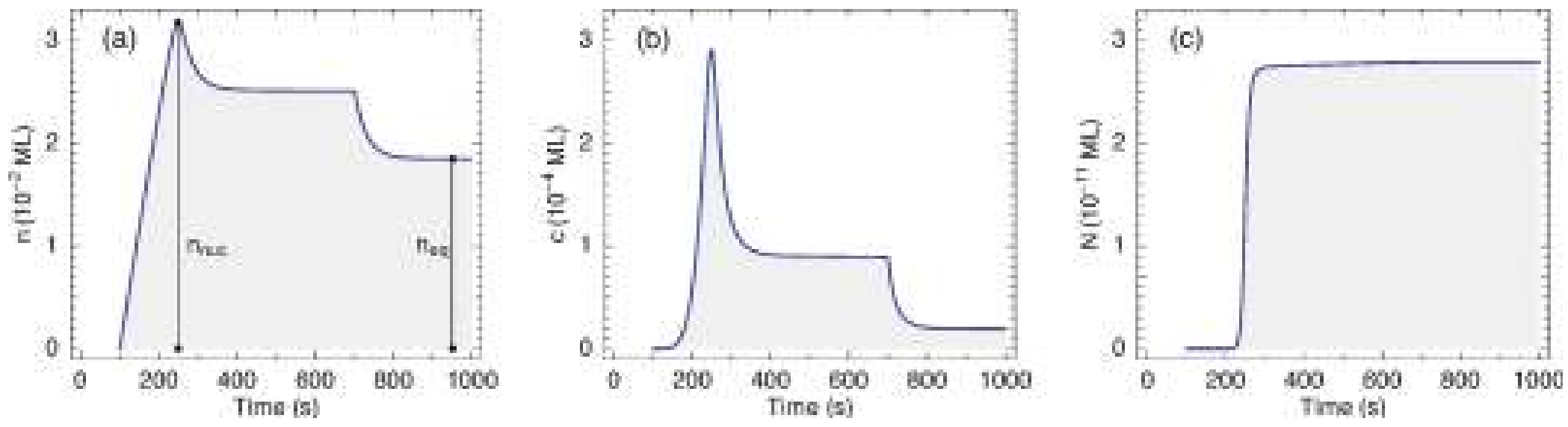

**Figure 4.** Time dependence of (a) the carbon adatom density n, (b) the five-atom mobile cluster density c and (c) the total immobile island density N obtained from the solution of the rate equations eqs 4-6 at T = 1020K. $n_{nuc}$ and $n_{eq}$ are indicated in (a). The kinetic parameters are $E_D$ = 0.92eV, $E_{D'}$ = 0.87eV, $E_K$ = 1.72eV, $E_{K'}$ = 1.27eV and j=6. From reference 17.

Figure 5 shows the solution for the carbon adatom density compared to the experimental data in Figure 5 of reference 15. We can see that the theoretical solution reproduces the main features of the Sandia group data. The rate equations account for the position of the nucleation peak at approximately 150 seconds after the carbon atom flux is turned on but underestimates the value by around 10%. The presence of a stationary adatom density with the flux turned on is accounted for but the rate equations overestimate the value by a few percent. Once the flux is turned off the value of the equilibrium densities for the rate equations and the experimental data are exactly the same.

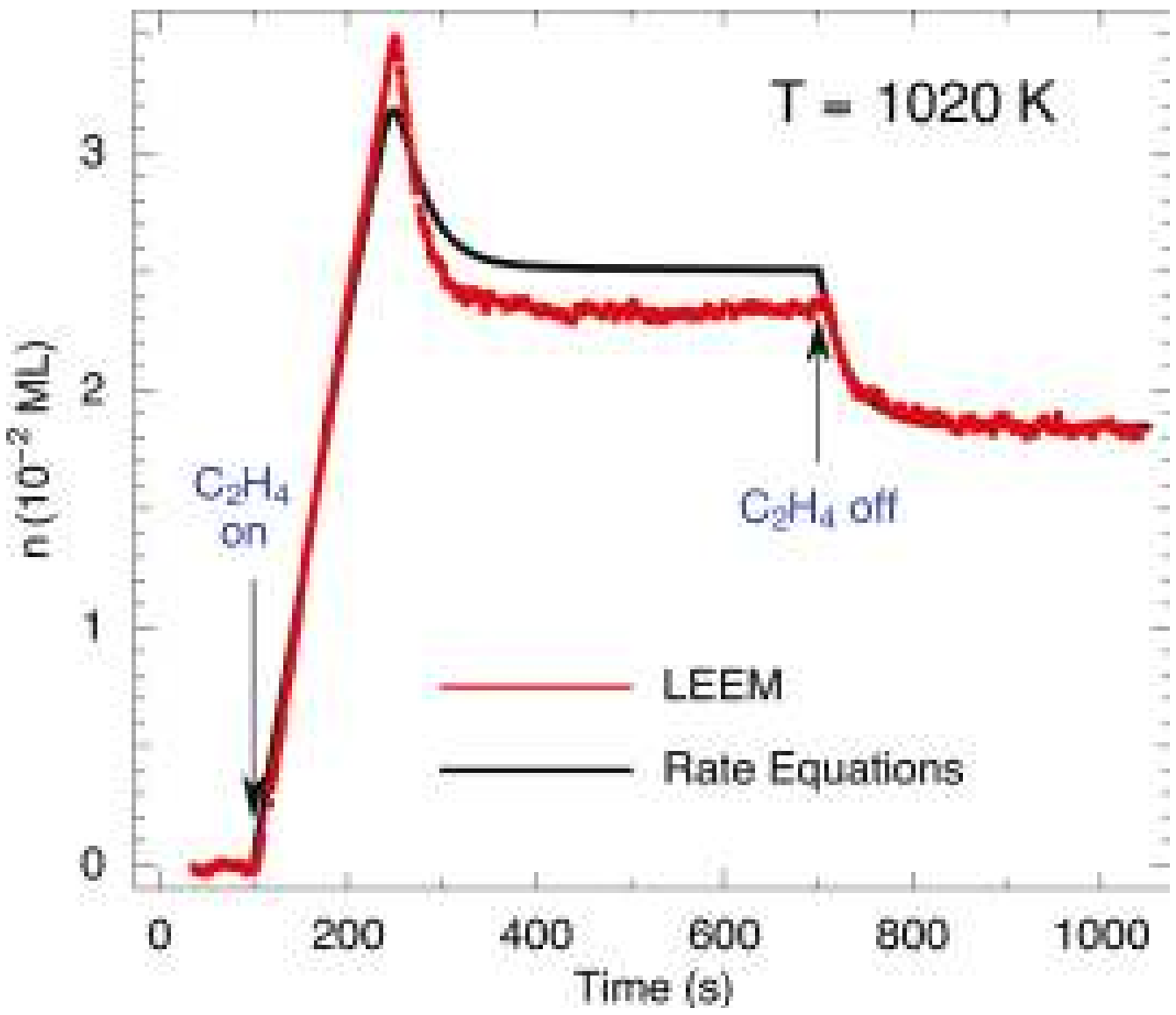

**Figure 5.** The evolution of the carbon adatom density in figure 4a (black curve) compared with the corresponding LEEM date of Loginova et al. (red data points). From reference 17.

We noted that the rate equations include no spatial information and it is likely that the lack of this refinement accounts for the discrepancies between the theoretical solution and the experimental data. The local spatial arrangements of islands, which is critical in establishing island growth rates,[20] is ignored and consequently could affect the value of the adatom density during the stationary regime with the carbon atom flux turned on. However $n_{eq}$ shows good agreement with the experimental data once the carbon atom flux is turned off. This may be because $n_{eq}$ is dominated by the detachment of adatoms from an island followed by re-attachment to the same island and so is relatively unaffected by the spatial arrangement of islands.

Figure 6 shows the temperature dependence of the nucleation and equilibrium densities of adatoms for both the rate equations and the experimental data. The choice of j=6 was required in order to give the best agreement for the nucleation density as a function of temperature. It is interesting to note that while the processes of single adatom attachment and detachment are not thought to significantly affect island growth they are included in the rate equations and, given the exact agreement between the rate equations and experimental data, seem to have observable consequences during equilibration.

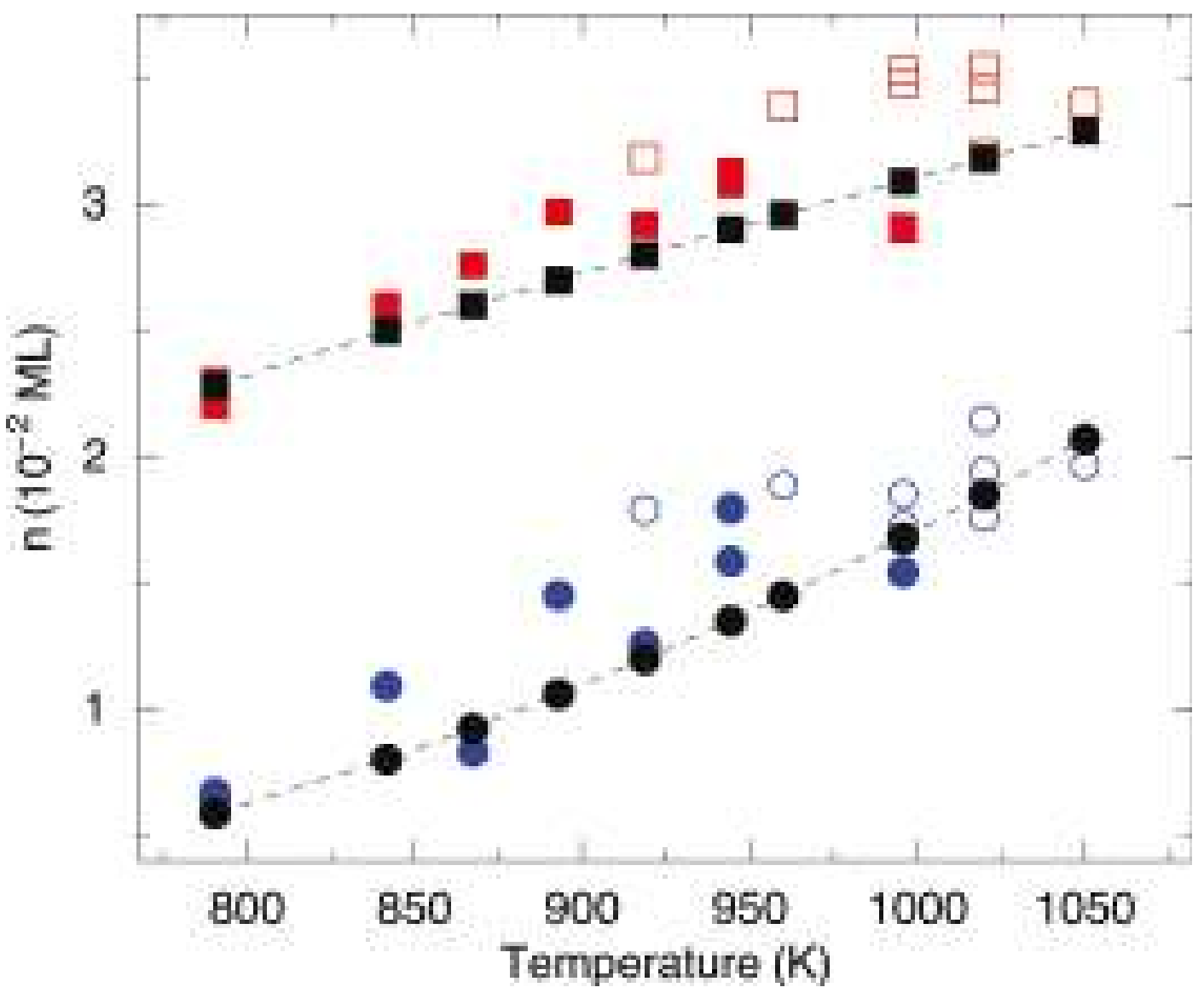

**Figure 6.** The temperature dependence of $n_{nuc}$ (red symbols) and $n_{eq}$ (blue symbols) compared to the corresponding quantities obtained from rate equations (black symbols). Experiments[15] were carried out with carbon vapour (filled symbols) and ethylene (open symbols). The dotted lines are a guide for the eye. From reference 17.

The work of Zangwill and Vvedensky suggests that the smallest islands that are direct precursors to graphene formation on metal surfaces are significantly large. Recent experimental work supports this conclusion. Cui and co-workers[21] deposited coronene ($C_{24}H_{12}$) on Ru(0001), the same substrate used by Loginova and co-workers,[14,15] and stepwise annealed the surface up to 1100K in ultrahigh vacuum (UHV). The coronene molecules dissociate above 550K and the uniform surface structures of around 1nm in size change to inhomogeneous islands with size varying from a few angstroms to a few nanometres. However uniform nanoclusters form on the surface again at higher temperatures, which approach the condition of graphene formation on Ru(0001).

Cui and co-workers observed two kinds of uniform nanoclusters on the surface annealed at 900K. The larger clusters have the same shape and size of coronene and are believed to be the same $7C_6$ structure ($C_6$ is a hexagonal ring of carbon atoms) as found in coronene but without the peripheral hydrogen atoms. In this case the edge carbon atoms interact strongly with the substrate causing the central carbon atoms to detach from the Ru surface which produces dome-shaped clusters. The smaller clusters found have a $3C_6$ structure and have size less than 0.6nm. Strong bonding between the peripheral carbon atoms and the substrate also results in dome-shaped structures for the smaller clusters.

Upon annealing the surface up to 1000K the smaller clusters become dominant and some graphene structures begin to form. At 1100K only graphene structures were observed on the surface. Therefore the carbon nanoclusters are the intermediate step in the growth of graphene sheets on the Ru(0001) surface. The $7C_6$ and $3C_6$ nanoclusters are the smallest graphene structures ever observed. We can make a comparison between these nanoclusters and the 30 atoms islands in the rate theory model of Zangwill and Vvedensky which are taken to be the smallest islands that lead directly to graphene formation.

Schaub and co-workers[22] have also observed graphene nanoclusters when investigating the nucleation and growth mechanisms of graphene on rhodium. They deposited ethene ($C_2H_4$) on the Rh(111) surface and then stepwise annealed the surface up to 973K in UHV. As the temperature increases the ethene rapidly dissociates to $C_2H_3$ and adsorbs in hexagonal close packed hollow sites on the surface. Further annealing up to 973K leads to the formation of uniformly sized carbon nanoclusters and coincides with the formation of graphene islands. The observed nanoclusters have a $7C_6$ structure.

Schaub and co-workers detected a significant decrease in nanocluster density as the temperature increased which they suggest means that the nanoclusters start to diffuse on the Rh(111) terraces and eventually coalesce to form graphene. This is a significant difference to the immobile 30 atom islands in the rate theory model inspired by the experimental data from the Sandia group. However the differences in the growth methodology and substrate used by Schaub and co-workers compared with those used by Loginova and co-workers may explain this difference. Both results do suggest that the formation of nanoclusters consisting of a significant number of carbon atoms is an intrinsic step in graphene formation.

## 4. Kinetic Monte Carlo simulations

The next step in seeking to understand the growth mechanisms of epitaxial graphene is to develop a model that includes spatial information. This could involve ‘unfolding’ the rate equations written down by Zangwill and Vvedensky[17] so as to include equations for the densities of islands of different sizes and then including island size dependent capture numbers that determine the propensity for adatoms and clusters to attach to islands of a given size.

This project involves carrying out Kinetic Monte Carlo (KMC) simulations of epitaxial growth involving mobile clusters of atoms that must collide to nucleate immobile islands. To the best of our knowledge, simulations of this kind have never been carried out before simply because there was no need to do so as no epitaxial systems other than Gr/Ru(0001) and Gr/Ir(111) have been shown to exhibit the nonlinear growth kinetics as described in section 2.

Out of equilibrium Monte Carlo simulations have become an important tool in surface science since they are an efficient way of simulating the movement of atoms on crystalline surfaces and explicitly take into account the spatial information and stochastic nature of thin film growth.[18,23] The observed motion of adatoms on a metal surface is of infrequent hops of atoms form one site on the surface to another adjacent site. For example, a single isolated atom on a (001) surface lies in a so-called four-fold hollow site in contact with four surface atoms simultaneously and will stay in the same site for comparatively long periods of time before making a quick hop to one of its four neighbouring sites.

When the adatom is in a four-fold hollow site it is in contact with the maximum possible number of surface atoms on a (001) surface. When it jumps from one site to another the atom has to pass through a succession of intermediate states where it is in contact with a smaller number of surface atoms. These intermediate states are therefore higher in energy than the preferred four-fold hollow sites and constitute an energy barrier to the hopping process. It is because of these energy barriers that adatoms stay in the same sites for relatively long periods of time.

The typical height, $E_{ij}$, of the energy barrier for hopping from a site i to a site j is significantly greater than the thermal energy $k_BT$ available for crossing it. The hopping rate, $r_{ij}$, at which an isolated adatom hops from site i to site j is given by the Arrhenius law:

$$r_{ij} = \nu_0 e^{-\beta E_{ij}} \tag{7}$$

where $\nu_0$ is the attempt frequency which sets the overall time-scale for the movement of adatoms and is given by $2k_BT/h$, in which $k_B$ is Boltzmann's constant, T is the absolute temperature and h is Planck's constant, and $\beta = 1/k_BT$. The form given here for the attempt frequency assumes that it is independent of adatom environment.

In order to simulate a system in which there are many adatoms on the surface we need to consider the effect on the energy barriers of the interactions between adatoms. The interactions affect both the binding energies of atoms in the four-fold hollow site and the energies of the intermediate states which atoms have to pass through when hopping from one site to another.

A model of adatom interactions that has been used to study general trends in surface science[18,24] is the nearest neighbour bond-counting model. In this model the energy barrier for an adatom with n nearest neighbours to hop is given by

$$E_{ij} = E_D + nE_N \tag{8}$$

where $E_D$ is the energy barrier for surface diffusion and $E_N$ is the nearest-neighbour bond energy.

In addition to the moves involving the hopping of adatoms the deposition of adatoms is included in the model. The rate of this process is given by a function F(t) which denotes the flux of adatoms arriving at the surface as a function of time.

The algorithm used to simulate epitaxial growth is known as the Kinetic Monte Carlo algorithm or N-Fold Way algorithm.[21,25] Considering the bond-counting model described above we can see that, since the hopping rate is entirely determined by the barrier height, there are z different hopping rates associated the system, where z is the lattice coordination number of the surface (4 in the case of the (001) surface that we have been considering). This is because n can only have integer values between 0 and z-1 inclusive – an adatom cannot hop to an adjacent site if it is surrounded by neighbouring adatoms.

The KMC algorithm applied to epitaxial growth involves initially dividing all possible moves on the lattice into z lists, one for each allowed value of n. All the moves in each list occur at the same rate:

$$r_n = \nu_0 e^{-\beta(E_D + nE_N)} \tag{9}$$

The KMC algorithm consists of four steps:

1. We choose a move at random from all the possibilities in proportion the hopping rate. We can achieve this by first picking one of the lists at random with probability proportional to $N_n r_n$ where $N_n$ is the number of moves in list n. Then we pick one of the moves from that list uniformly at random.

2. We perform the chosen move.

3. We add an amount Δt to the time, where Δt is given by

$$\Delta t = \frac{-\ln(\rho)}{\sum_n N_n r_n} \tag{10}$$

   where ρ is a random number distributed uniformly over the range (0,1).

4. We update the lists of moves to take the move we have just made into account.

In the last step we do not need to recalculate all the lists of moves from scratch for the entire lattice since most sites on a large lattice will be far enough away not to be affected by the move. However, if an adatom has hopped from site i to site j, the hopping rates for all moves to or from nearest-neighbour sites of either site i or site j will be affected and the corresponding entries need to be found in the lists and moved to different lists. There will also be some new moves into the site vacated by the adatom which just hopped that become possible and some other moves into the site the adatom now occupies that are no longer possible. These moves will also need to be added or subtracted from the lists. The advantage of the KMC algorithm is that a move is definitely executed at each Monte Carlo step. This improvement in efficiency makes up for the additional computing time required for updating the lists of moves after each step.

The question arises as to how we can be sure that the code we write is correctly simulating the model of epitaxial growth by cluster attachment. We are fortunate in that we can use the rate equations developed by Zangwill and Vvedensky[17] as a consistency check. If the results of the simulations do not at all resemble the results from the rate equations then it is likely that there is a problem with the code. As a further measure to ensure the correct implementation of the simulations two completely independent versions of the code were developed. We checked the results of the separate simulations for consistency and if discrepancies were found then we checked both versions of the code for errors and made the appropriate changes so that the results matched.

The technique first presented by Hoshen and Kopelman in 1976 was used to count the number and size of islands during our simulations. The details of the algorithm can be found in reference 26.

## 5. A simple model of epitaxial growth by adatom attachment

Before we move on to consider simulations of epitaxial growth by cluster attachment we first present the results of a KMC simulation of a simple model of epitaxial growth involving only nearest-neighbour interactions between adatoms on a square lattice. This so called pair bond solid-on-solid model has been used for many years by the surface science community to investigate the nucleation and growth of two dimensional islands in epitaxial growth.[18,24] The results of these simulations will help to inform our analysis of a model of a non typical epitaxial system involving mobile clusters.

The processes involved in the simple model are as follows: atoms are deposited onto the surface and these adatoms hop around the lattice with an energy barrier that depends on the number of nearest-neighbours as described in the section 4. The simulations were carried out on a square lattice of 200x200 sites. 0.10 monolayers (ML) of atoms were deposited over 0.10 seconds with flux 1.0MLs$^{-1}$ and the system was then allowed to relax until reaching equilibrium.

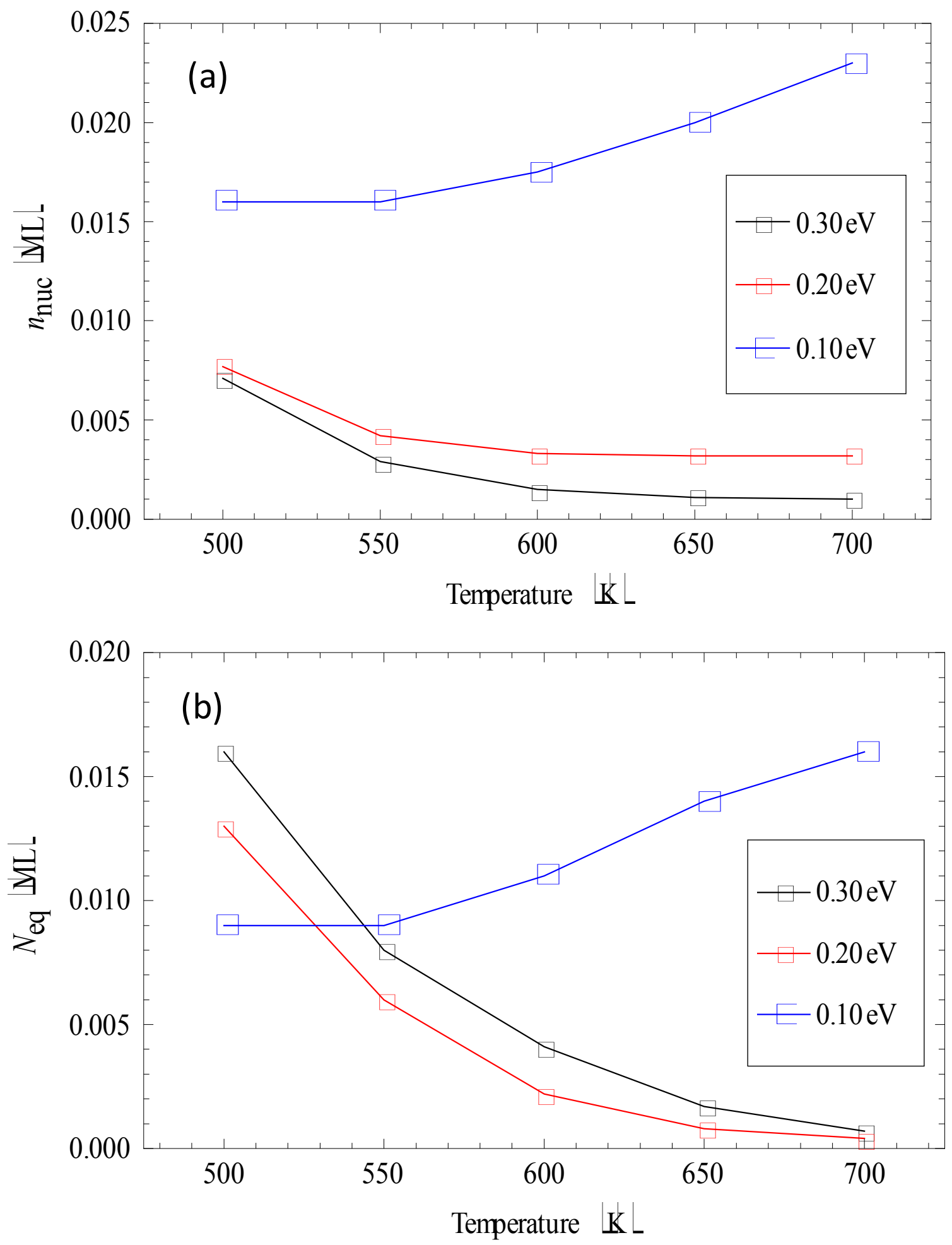

**Figure 7.** Temperature dependence of (a) the nucleation density of adatoms, $n_{nuc}$, and (b) the equilibrium density of islands, $N_{eq}$. Kinetic parameters are $E_D$ = 1.00eV with $E_N$ as given in the plot legends. 0.10ML coverage was deposited at a flux of 1.0MLs$^{-1}$ and the system was then allowed to relax until equilibrium was reached.

Figure 7a shows the temperature dependence of the adatom nucleation density, $n_{nuc}$, and figure 7b shows the temperature dependence of the equilibrium island density, $N_{eq}$, for three different nearest-neighbour bond energies; 0.10eV, 0.20eV and 0.30eV. Islands are counted as any group of two or more neighbouring atoms. The adatom nucleation density is the peak value of the density of adatoms. At this density a stable island (or islands) has formed and begins to grow by capturing further adatoms which causes the adatom density to decrease from its maximum value.

For bond energies of 0.20eV and 0.30eV the nucleation adatom density decreases with increasing temperature. This behaviour suggests that the critical island size, the size of the smallest islands that do not dissociate, is approximately independent of temperature. Therefore $n_{nuc}$ decreases because the rate of adatom diffusion increases with temperature and so adatoms explore the surface more rapidly and hence require a smaller density to form a stable island.

For nearest-neighbour bond energy of 0.10eV the nucleation adatom density increases as temperature increases. We suggest that this is because with such a low bond energy the rate of adatom detachment is appreciable in comparison to the rate of adatom diffusion and causes the critical island size to increase with increasing temperature. Hence the effect of a faster rate of adatom diffusion with increasing temperature is negated and a greater density of adatoms is required to form a stable island.

We can see that for bond energies of 0.20eV and 0.30eV the equilibrium island density decreases with increasing temperature. As the rate of diffusion of adatoms increases with increasing temperature islands grow more rapidly by adatom attachment. It therefore becomes increasingly difficult for new islands to nucleate as they will lose out to larger pre-existing islands in the competition to capture adatoms.

However, for nearest-neighbour bond energy 0.10eV we can see that the equilibrium island density increases as a function of temperature. This is because at higher temperatures stable islands find it increasingly difficult to form with 0.10ML coverage and a large fraction of the island density is due to small islands that are present on the lattice but are constantly nucleating and breaking up. As the system temperature increases, the number of stable islands decreases but the number of smaller transient islands increases. The overall effect is that the total island density increases with increasing temperature.

Another quantity we can measure for this model is the island size distribution. As first suggested on the basis of simulation results,[27] this quantity takes the form,

$$n_s = \frac{\theta}{(s_{av})^2} g(s / s_{av}) \tag{11}$$

where $n_s$ is the density of islands of size s, θ is the coverage, g(x) is a scaling function and $s_{av}$ is the average island size given by,

$$s_{av} = \frac{\sum_s s n_s}{\sum_s n_s} = \frac{\sum_s s n_s}{N} \tag{12}$$

That is to say that if you were to pick an island at random on the lattice then $s_{av}$ is the average size of the island that you would choose. An alternative definition of the average island size is sometimes used in the literature, for example in reference 19.

The scaling function g(x) is usually taken to be indexed by the critical island size.[24] This means that for systems in which the critical island size is independent of temperature a plot of $n_s(s_{av})^2 / \theta$ against $s / s_{av}$ for different temperatures should show data collapse. Figure 8 shows the scaled island size distributions for $E_N$ = 0.30eV for temperatures in the range 500-700K.

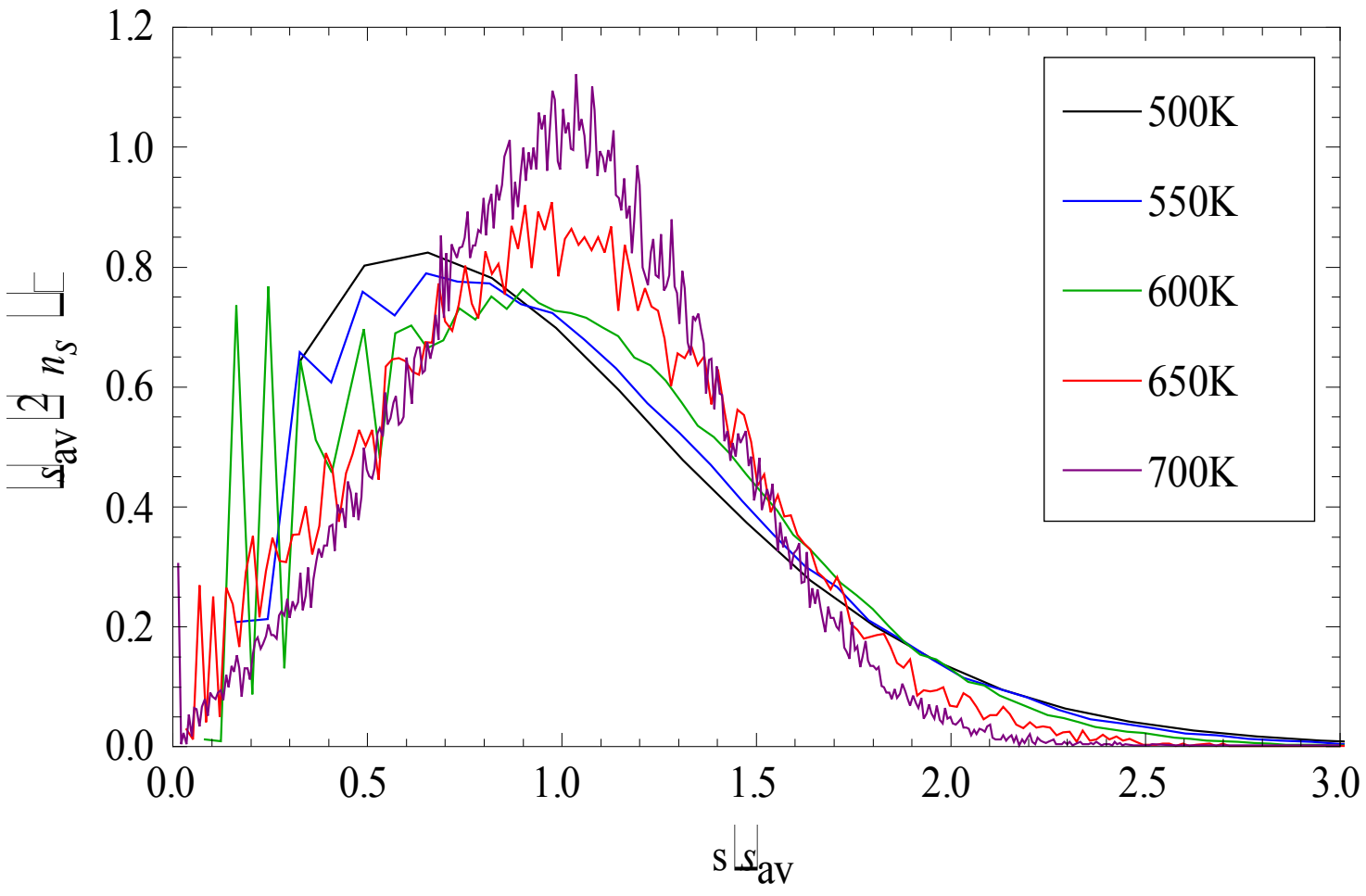

**Figure 8.** Scaled island size distributions for the pair bond solid-on-solid model. Kinetic parameters are $E_D$ = 1.00eV and $E_N$ = 0.30eV. Temperatures are given in the plot legend.

The distribution gets sharper around s / $s_{av}$ as the temperature increases. This is because fewer islands nucleate at higher temperatures and so there is less variation in island size around the average value. The absence of perfect data collapse is indicative that the critical island size for this system has some weak dependence on temperature. In contrast we can see in figure 9 that the island size distributions do exhibit data collapse for the case of irreversible aggregation when $E_N$ is infinite and islands have zero probability of dissociating.

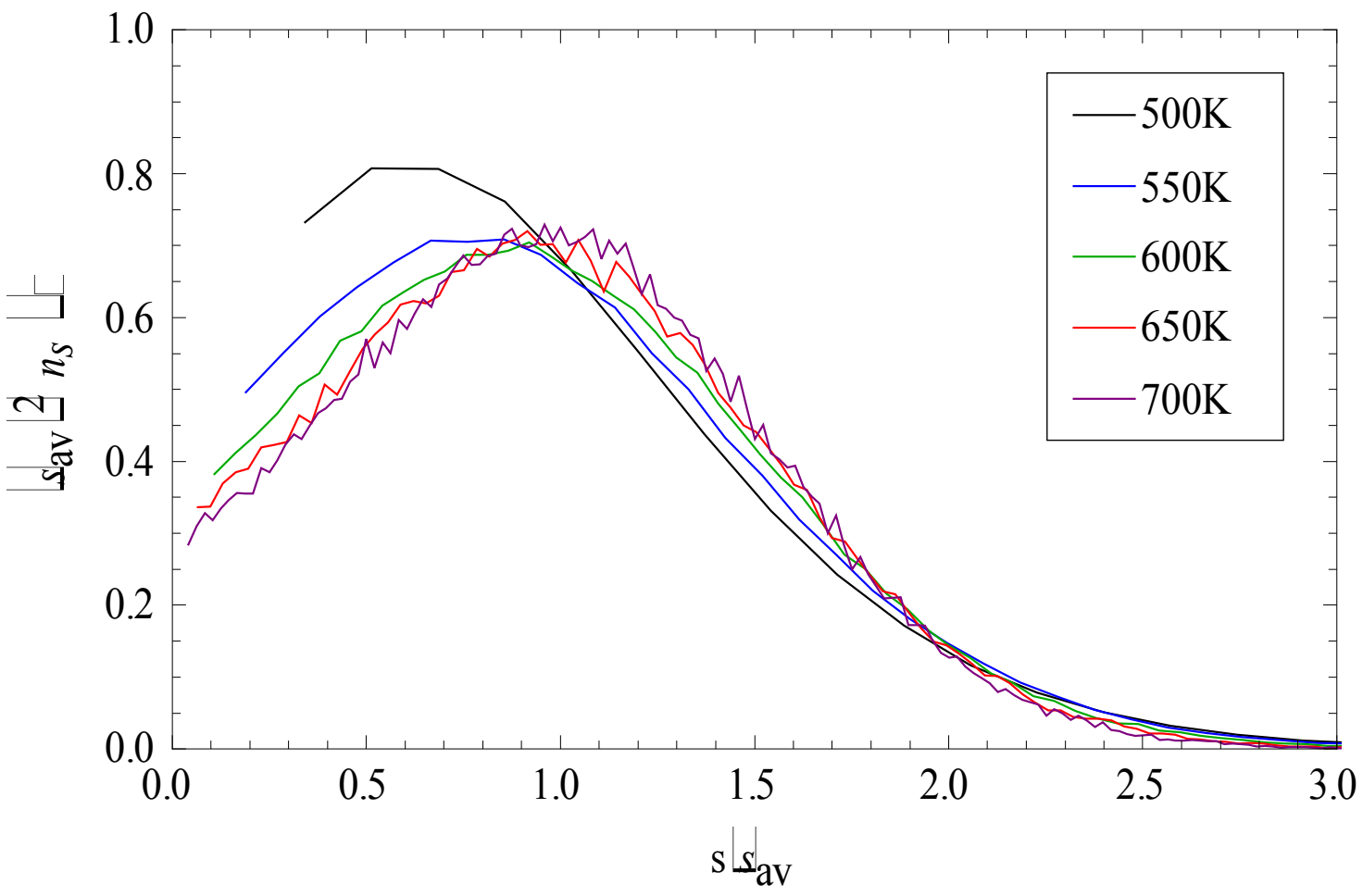

**Figure 9.** Scaled island size distributions for irreversible island growth. Kinetic parameters are $E_D$ = 1.00eV. Temperatures are given in the plot legend.

# 6. A model of epitaxial growth by cluster attachment

We now move on to consider a model of epitaxial growth involving mobile clusters of atoms. We do not seek to replicate the results of the Sandia group because the high temperature and low carbon atom deposition rate used in the LEEM experiments render KMC simulations impractical. However we do aim to qualitatively replicate some of the novel features of the Gr/Ru(0001) system; in particular the temperature dependence of the density of adatoms at island nucleation, the relationship between the values of the nucleation density and equilibrium density of adatoms and the sharp decrease in island density with increasing temperature

We consider a growth model on a square lattice in which the mobile clusters involved in island growth are tetramers made up of four atoms. These tetramers have spatial extent and still occupy four sites on the lattice. For simplicity we only allow a 2x2 arrangement of atoms to bind together to form a cluster. A 2x2 arrangement is also the only configuration in which all four atoms making up the cluster have two bonds to other atoms in the cluster and so we can make a physical argument that this is the most stable configuration. These clusters are mobile and have an energy barrier for diffusion. In seeking to develop a model that has some similarities to the graphene system we require that the energy barrier for tetramer diffusion is lower than the energy barrier for adatom diffusion. In the graphene system this arises because strong carbon-carbon bonds within a five-atom cluster weaken the barrier for the cluster to diffuse over the metal substrate. The tetramers are considered a distinct species in the system and when they hop across the lattice the constituent atoms move as one and the cluster retains its original configuration. There is no interaction between adjacent single adatoms and tetramers.

Since we require four atoms to make a cluster the possibility arises for two or more adatoms to occupy neighbouring sites but not form a cluster. In our model these adatoms are not affected by the presence of their neighbours and there is no intermediate bonding between them. They still behave like free adatoms with zero nearest-neighbour bonds but cannot hop into a site occupied by a neighbouring atom. Although perhaps not being completely physically realistic this feature helps to keep the model as simple as possible.

If we were to consider some form of bonding between neighbouring adatoms before they formed a cluster we would have to decide whether these structures were mobile or immobile. If these intermediate species were mobile then a situation could arise in which two groups of three atoms came together – would four of the atoms form a cluster and the remaining two form another intermediate two atom structure? Other similar scenarios would also need to be accounted for in the simulation. In seeking to develop a simple model of epitaxial growth by cluster attachment we ignore these complications and so consider only three species in our model: adatoms that are only bonded to the surface, mobile clusters of four atoms in a 2x2 configuration and atoms with one or more nearest-neighbour bonds that make up immobile islands.

Mobile clusters are not infinitely long lived in the model but have an energy barrier for breaking up into four adatoms. The rate at which they break up into four adatoms is determined by the energy barrier for the process. In our model we constrain the energy barrier for dimer break up to be larger than the energy barrier for dimer diffusion. When a cluster breaks up into four adatoms one of the atoms making up the cluster hops away from it. The remaining three atoms then behave as described above and do not feel the presence of their neighbours.

When a certain number of clusters come together they form an immobile island so we are effectively specifying the critical island size of the system. We only consider certain configurations of tetramers to form islands. In our simulations we consider the situations when two, four and six clusters are required to come together to form an island. Table 1 below show the configurations of clusters on the square lattice for each of the three cases that we allow to form an immobile island.

| Number of tetramers required to form an immobile island | Allowed configurations of tetramers that form an immobile island | Critical island size (atoms) |
|---|---|---|
| 2 | 1x2<br>2x1 | 8 |
| 4 | 2x2 | 16 |
| 6 | 2x3<br>3x2 | 24 |

**Table 1.** Configurations of tetramers that form an immobile island for different values of j, the number of tetramers required to form an immobile island.

Once an immobile island has formed we no longer consider it to be made up of tetramers but simply to be made out of atoms. The atoms in the nucleated island 'forget' that they were previously part of mobile clusters and their energy barrier to hopping is now determined by the number of nearest-neighbour bonds they have, as in the simple model described in section 5. Islands grow by capturing mobile clusters and adatoms. We only allow an island to capture a tetramer if two of the atoms making up the tetramer form a bond to the island edge. Once the tetramer has joined the island its constituent atoms once again behave as individual atoms whose energy barrier to hopping is determined by their number of nearest-neighbour bonds.

There is no intermediate bonding between tetramers when fewer tetramers than the number required to nucleate an island meet on the lattice. The tetramers feel no effect from their neighbours other than a restriction of the sites to which they can hop.

The five parameters in the model described above are summarised below:

- $E_D$ - the energy barrier for surface diffusion of adatoms
- $E_{D'}$ - the energy barrier for surface diffusion of tetramers
- $E_K$ - the energy barrier for tetramers breaking up into four adatoms
- $E_N$ - the bond energy between atoms forming an immobile island
- j - the number of tetramers that must collide to form an immobile island

# 7. Tetramer Model Results

## 7.1 Comparison between rate equations and simulations

The simulations of the tetramer model were carried out on a square lattice of 200x200 sites. Figures 10a, 10b and 10c show a comparison between the results of our KMC simulations for the tetramer model with the given set of parameters compared to the appropriate rate equations. The parameters D, D’ and K in the rate equations have direct equivalents in the simulations but the simulations use a nearest-neighbour bond energy rather than the single detachment parameter K’ used in the rate equations. Therefore the parameter K’ in the rate equations was tuned to give the best agreement between the simulations and the rate equations for the position and the value of the adatom nucleation peak. Figure 10a shows that the adatom density has a similar temporal profile to the one seen from the graphene experiments. Initially the density increases linearly with time before reaching a peak value when island nucleation occurs. It then decreases until reaching a stationary value while the adatom flux is still turned on. After the adatom flux is switched off the density decreases further until adatoms reach equilibrium with mobile clusters and islands.

We can see that although there is good agreement between the rate equations and KMC simulations on the position and value of $n_{nuc}$, the adatom density in the simulations decreases far more rapidly and has a much lower stationary value while adatoms are still being deposited on the surface. We put this down to the inclusion of spatial effects in the simulations; as the islands have spatial extent they capture adatoms more effectively than the point islands represented by the rate equations. We can also see that in the simulations the value of $n_{eq}$ is more than an order of magnitude smaller than the value of $n_{nuc}$.

There is good agreement between the rate equations and simulations for the position of the peak value of the density of tetramers although the rate equation value is greater than the simulation value by around 25%. The rate equations also significantly underestimate the rate at which the tetramer density decreases after island nucleation; in the simulations the tetramer density decreases rapidly to zero after nucleation. Again we put this down to the effect of the spatial information included in the simulations and the islands capturing tetramers more effectively when they have spatial extent.

The rate equations significantly underestimate the value of the island density at equilibrium with adatoms and clusters given by the KMC simulations. This result is not surprising as the rate equations for the graphene system give a miniscule island density of order $10^{-11}$ML which is not in agreement with experiment. We can also relate this back to the rapid decrease in adatom and cluster densities given by the simulations; the greater the number of islands present in the system, the greater the rate of attachment of adatoms and clusters to islands.

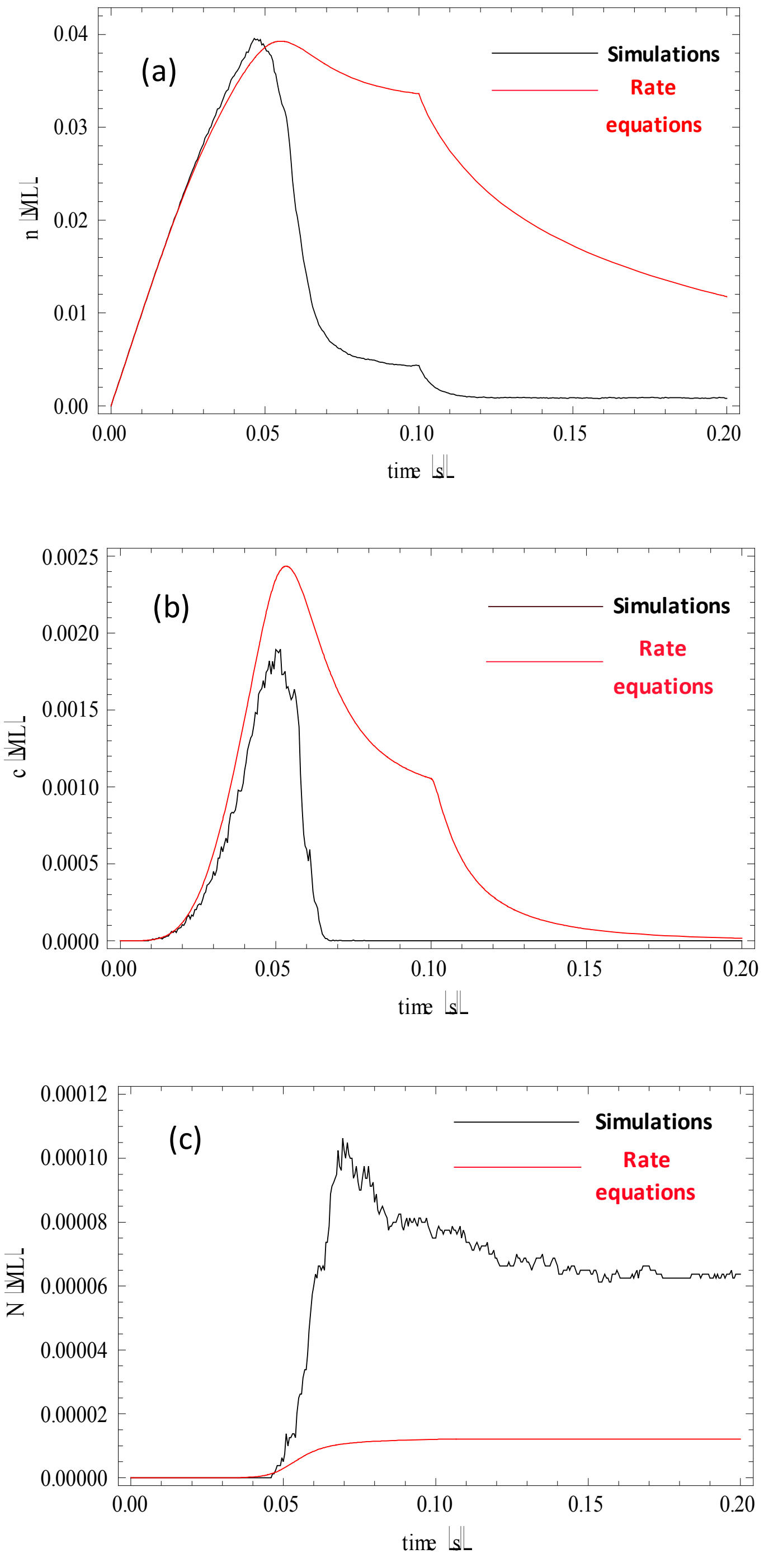

**Figure 10.** Time dependence of (a) the adatom density n, (b) the tetramer density c and (c) the island density N for j =4 at 650K. The black curves are the simulation results and the red curves are the results of the appropriate rate equations. Kinetic parameters are $E_D$ = 1.00eV, $E_{D'}$ = 0.80eV and $E_K$ = 1.40eV. $E_N$ = 0.20eV in the simulations. $E_{K'}$ = 1.60eV in the rate equations. 0.10ML coverage was deposited at a flux of 1.0MLs$^{-1}$ and the system was then allowed to relax for 0.10s.

There is however reasonable agreement between the two method for the time period over which most island nucleation occurs; between 0.05 and 0.07 seconds. We also note the slightly curious behaviour of the island density given by the simulations after the initial nucleation; it decreases by about 25% from its peak value while reaching equilibrium. It seems as though some smaller islands lose out in the competition for adatoms and clusters and break up during the equilibration process.

## 7.2 Temperature dependence of $n_{nuc}$

One of the features of the graphene system which we wish to account for qualitatively is the temperature dependence of the adatom density at the onset of island nucleation. Figures 11a, 11b and 11c show $n_{nuc}$ as a function of temperature with varying energy barriers for cluster break up, for the situations when two, four and six tetramers are required to collide to form an island respectively (j=2, j=4 and j=6). $E_N$ was set to 0.20eV which does not produce increasing $n_{nuc}$ with temperature in the growth model involving only mobile adatoms.

We can see that when j=2 $n_{nuc}$ decreases with increasing temperature for the range of cluster break up energies shown. When j=4 $n_{nuc}$ increases with increasing temperature in the range 600-700K for cluster break up energy 1.20eV. $n_{nuc}$ also increases between temperatures of 650-700K for cluster break up energy 1.40eV. When j=6 $n_{nuc}$ increases across the entire temperature range for cluster break up energies of 1.20eV and 1.40eV. It seems that in order to obtain the correct temperature dependence for $n_{nuc}$ we require both that a relatively large number of clusters are required to come together to form an island and that the rate of cluster break up is high enough to have an appreciable effect across the temperature range considered. In order for immobile islands to form in the system the probability of j clusters colliding must become large enough for the first island to nucleate. The probability of a j cluster collision occurring in a given time interval depends on both the density of clusters and the rate of cluster diffusion and is clearly an increasing function of both the cluster density and the diffusion rate. In this discussion the energy barrier for cluster diffusion is fixed and hence the rate of cluster diffusion depends only on the temperature of the system.

There are two competing processes that determine the temperature dependence of the density of adatoms required to nucleate an island. As the temperature increases and the rate of cluster diffusion increases there is a corresponding decrease in the density of clusters required for immobile island nucleation to occur. Plots 12a, 12b and 12c shows that for j=2, j=4 and j=6 the density of clusters required for island nucleation decreases with increasing temperature. As the density of clusters required to nucleate an island decreases the adatom nucleation density decreases.

If the energy required for cluster break up is prohibitively large then once clusters are formed they are effectively infinitely long lived in the system. If clusters have zero probability of breaking up then the density of adatoms required to form the necessary number of clusters for island nucleation to occur also decreases with temperature and therefore $n_{nuc}$ decreases as a function of temperature.

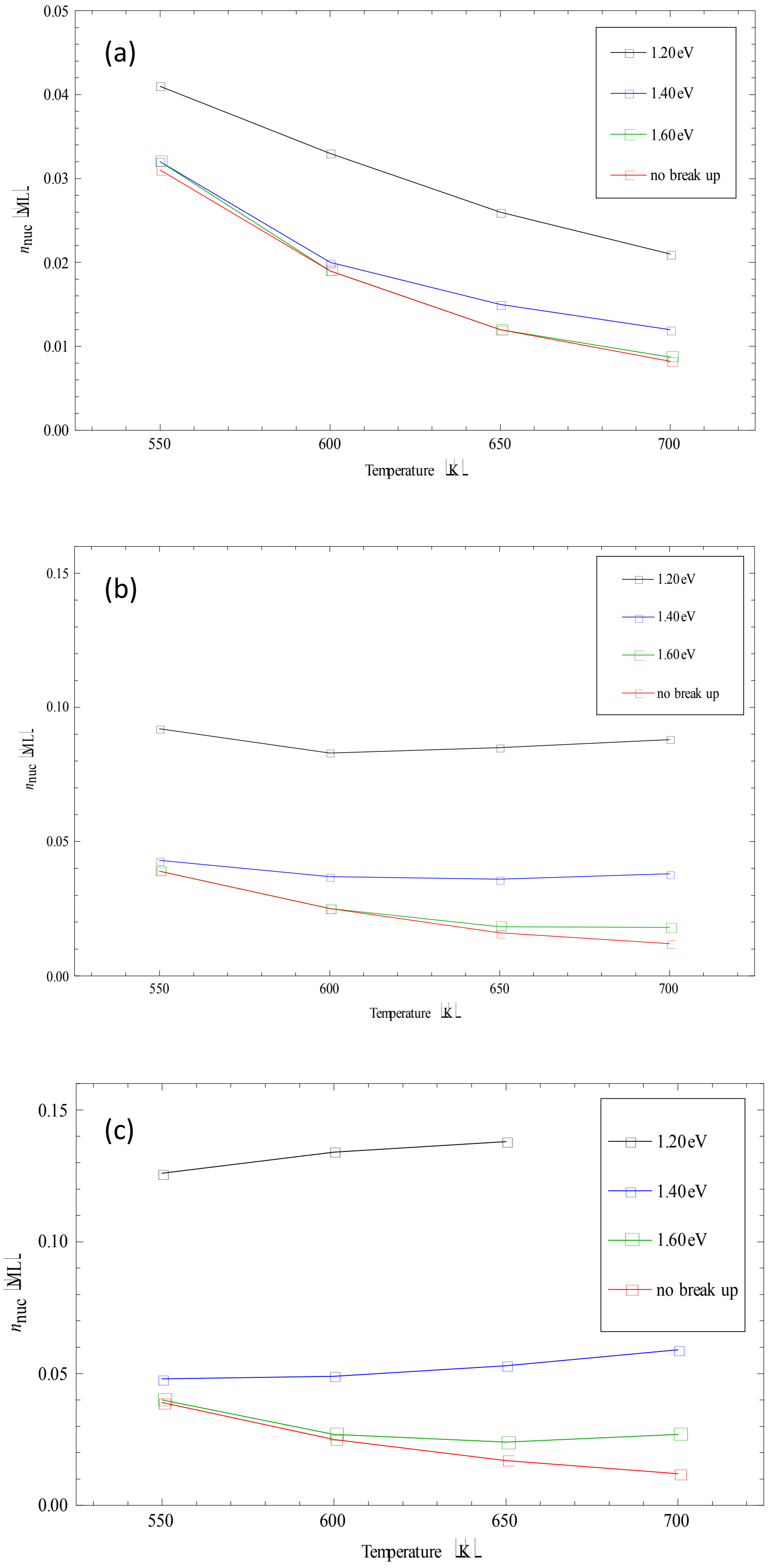

**Figure 11.** The density of adatoms at island nucleation, $n_{nuc}$, as a function of temperature for (a) j=2, (b) j=4 and (c) j=6. Kinetic parameters are $E_D$ = 1.00eV, $E_{D'}$ = 0.80eV and $E_N$ = 0.20eV. The values of $E_K$ are given in the plot legend. 0.10ML coverage was deposited at a flux of 1.0MLs$^{-1}$ for the j=2 and j=4 systems. 0.20ML coverage was deposited at a flux of 1.0MLs$^{-1}$ for the j=6 system. The 700K value for $E_K$ = 1.20eV is not shown because the simulations took too long to obtain reasonable data.

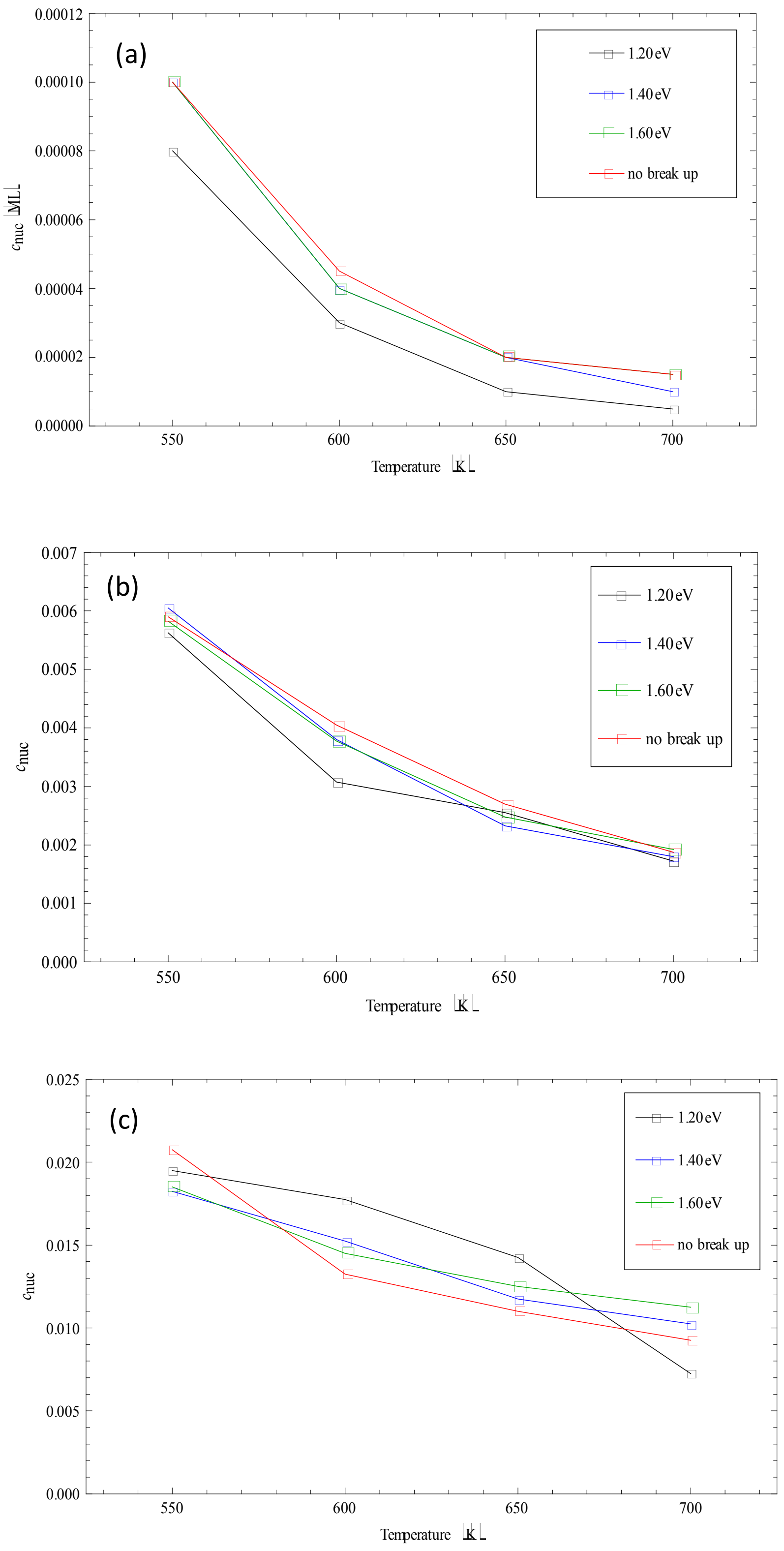

**Figure 12.** The density of clusters at island nucleation, $c_{nuc}$, as a function of temperature for (a) j=2, (b) j=4 and (c) j=6. Kinetic parameters are $E_D$ = 1.00eV, $E_{D'}$ = 0.80eV and $E_N$ = 0.20eV. The values of $E_K$ are given in the plot legend. 0.10ML coverage was deposited at a flux of 1.0MLs$^{-1}$ for the j=2 and j=4 systems. 0.20ML coverage was deposited at a flux of 1.0MLs$^{-1}$ for the j=6 system.

If the energy barrier for cluster break up is low enough then the density of clusters that can coexist with a fixed density of adatoms decreases as the temperature increases. Therefore as the system temperature increases the increased rate of cluster break up means that the density of adatoms required to reach a critical density of clusters increases.

For the adatom nucleation density to increase as a function of temperature the rate of cluster break up must be the most important process in determining the adatom nucleation density. Our results show that this happens when the number of clusters required to collide to nucleate an island is sufficiently high. The j=6 system is most affected by the decreasing cluster break up energy and shows the $n_{nuc}$ temperature dependence that we are looking for. The dependence of $n_{nuc}$ on both the cluster break up energy and the number of tetramers required to form an immobile island illustrates the increased complexity of the epitaxial system when it includes mobile clusters.

## 7.3 Relative values of $n_{nuc}$ and $n_{eq}$

From the comparison of the rate equations and KMC simulations we saw that the simulations typically give a value of $n_{eq}$ is more than an order of magnitude smaller than the value of $n_{nuc}$. This is a significant difference to the graphene system where Loginova and co-workers reported that $n_{nuc} \approx 2n_{eq}$. The explanation for the growth of graphene by carbon cluster attachment was that carbon adatoms have a large energy barrier to attach to graphene. In the model of epitaxial growth by cluster attachment described in section 6 there is no such barrier; once immobile islands have formed from a j cluster collision then adatoms are captured by islands as soon as they occupy a lattice site next to the island edge.

We introduce a modification to the model in the form of an energy barrier for adatom attachment to islands. When an adatom hops into a lattice site adjacent to the island edge it now attaches to the island with probability given by,

$$e^{-E_{ATOM}/k_B T} \qquad (13)$$

where $E_{ATOM}$ is the energy barrier for adatom attachment. This is not the most accurate method for simulating the dynamics of adatoms in which the energy barriers depend on both the initial and final state configurations of each move but we simply did not have time to carry out more complex modifications of the model.

Figure 13 shows the temperature dependence of the ratio of $n_{nuc}$ to $n_{eq}$ for a number of different values of $E_{ATOM}$ for simulations with j=6. We can see that as the energy barrier to adatom attachment increases the ratio of $n_{nuc}$ to $n_{eq}$ decreases. This behaviour is expected as introducing an energy barrier to adatom attachment does not affect the adatom density required to nucleate islands but will cause an increase in the equilibrium density of adatoms. Both values of $E_{ATOM}$ = 0.30eV and $E_{ATOM}$ = 0.40eV give $n_{nuc} \approx 2n_{eq}$ across the temperature range shown, as seen in the graphene system. We do not claim that this result is of quantitative relevance to the graphene system but we can conclude that a significant energy barrier for adatom attachment to islands is required in our model for there to be a substantial density of adatoms at equilibrium.

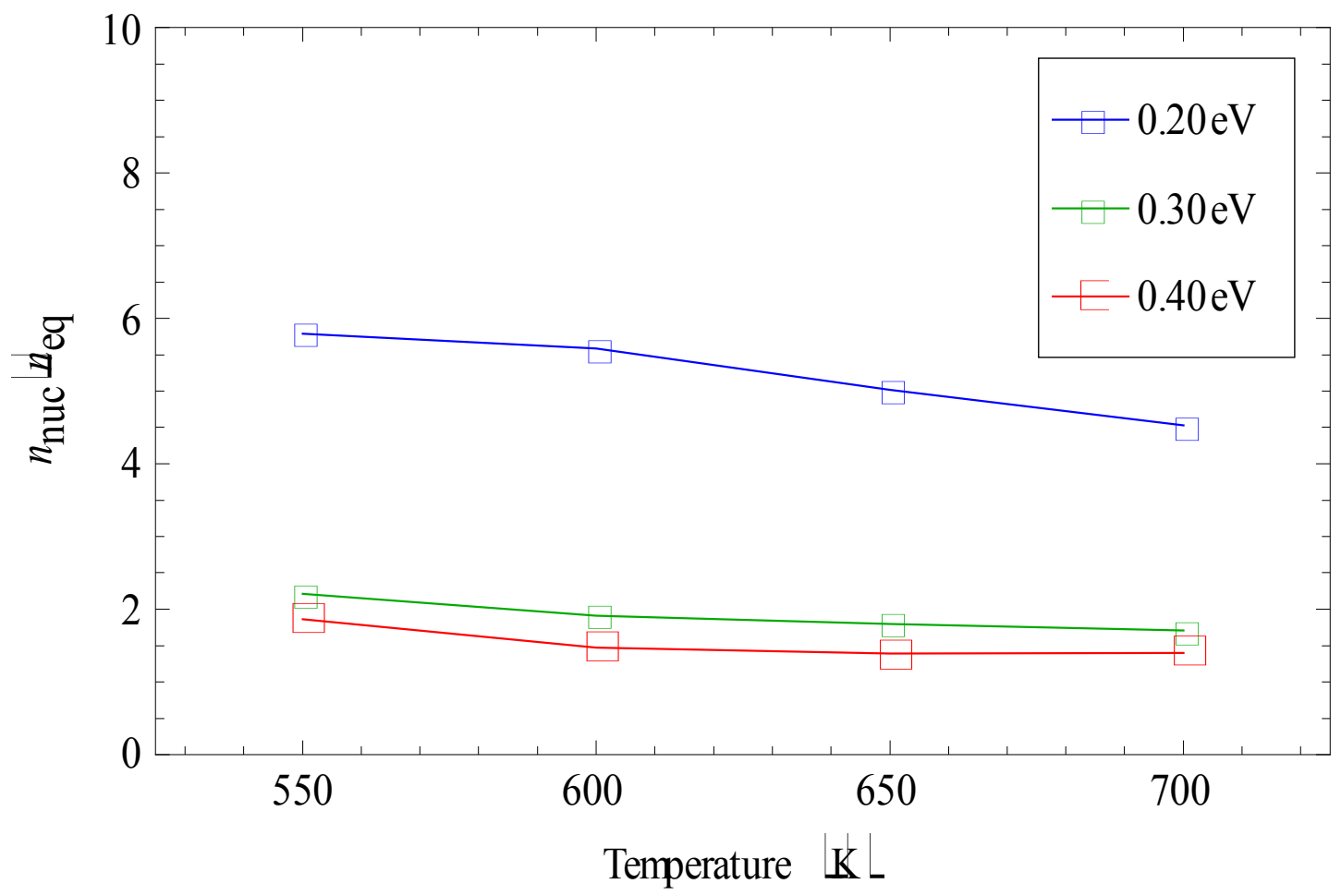

**Figure 13.** The ratio of $n_{nuc}$ to $n_{eq}$ as a function of temperature for j=6. Kinetic parameters are $E_D$ = 1.00eV, $E_{D'}$ = 0.80eV, $E_K$ = 1.40eV and $E_N$ = 0.20eV. The values of $E_{ATOM}$ are given in the plot legend. 0.20ML coverage was deposited at a flux of 1.0MLs$^{-1}$ and the system was then allowed to relax until equilibrium was reached.

## 7.4 Temperature dependence of the island density

The final feature of the epitaxial graphene system that we wish to qualitatively account for with our model is the result reported by the Sandia group that the nucleated island density dramatically decreases with increasing temperature. Since we have introduced an energy barrier for adatom attachment to immobile islands we also can consider an energy barrier for tetramer attachment to islands. This further modification of the model means that when a tetramer hops into a position on the lattice where it has two of its constituent atoms adjacent to the island edge it attaches with a probability given by,

$$e^{-E_{CLUSTER}/k_B T} \qquad (14)$$

where $E_{CLUSTER}$ is the energy barrier for tetramer attachment. We require that $E_{CLUSTER} < E_{ATOM}$ since the clusters are more weakly bound to the substrate than single adatoms.

Figure 14 shows the temperature dependence of the island density at equilibrium for a number of different values of $E_{CLUSTER}$ with j=4 and $E_{ATOM}$ = 0.30eV. Since we are considering the island density and not the adatom density we carried out these simulations with j=4 in order to reduce the computing time required. We can see that the higher the energy barrier for cluster attachment the greater the value of the island density at low temperatures. For $E_{CLUSTER}$ = 0.15eV, 0.20eV and 0.25eV the density decreases to a value of 2.5x10$^{-5}$ML at 700K. This corresponds to a single island on a 200x200 lattice. However, the behaviour for $E_{CLUSTER}$ = 0.05eV and 0.10eV is different and we will discuss it shortly.

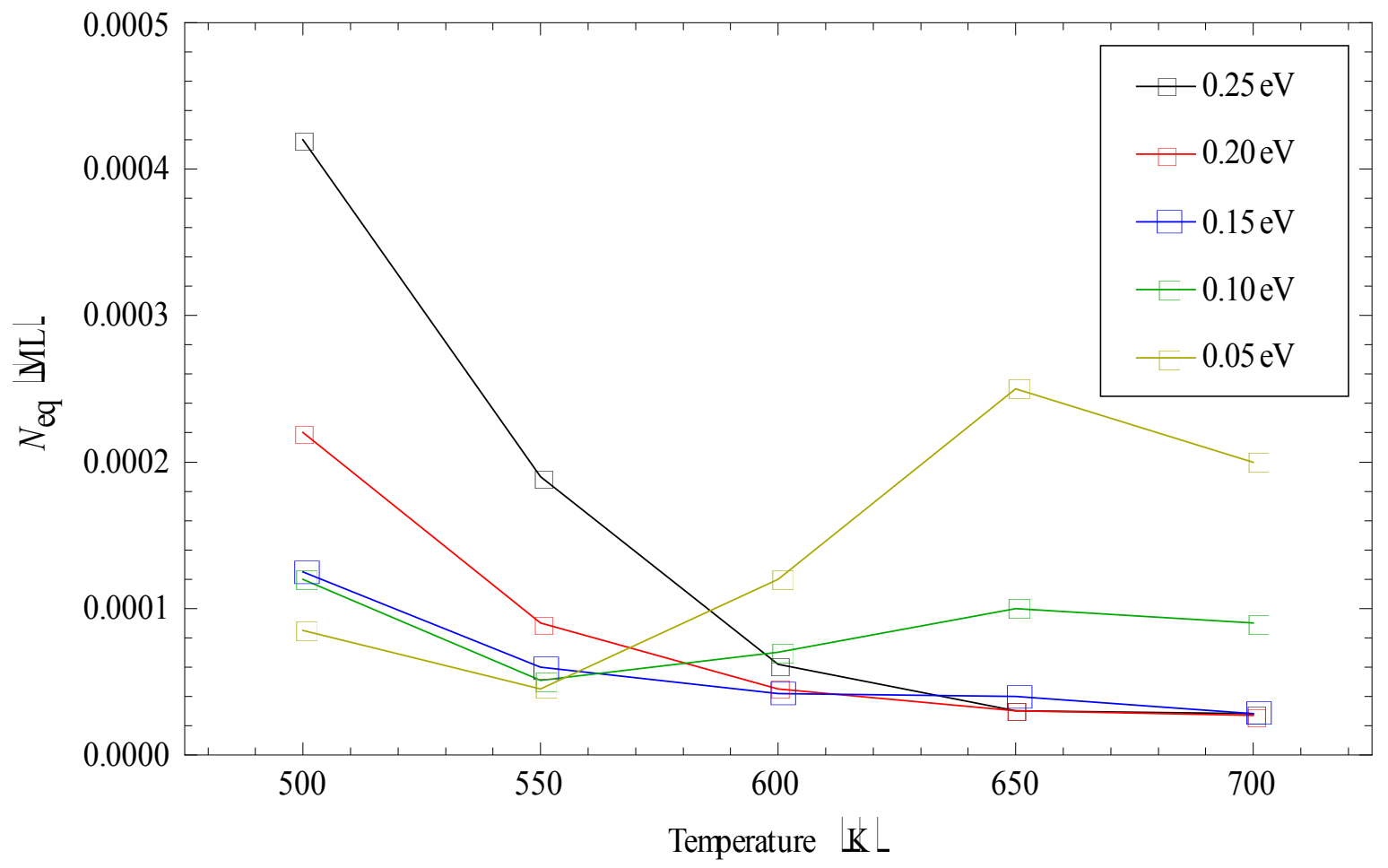

**Figure 14.** The temperature dependence of the equilibrium density of immobile islands, $N_{eq}$, for j=4. Kinetic parameters are $E_D$ = 1.00eV, $E_{D'}$ = 0.80eV, $E_K$ = 1.40eV, $E_N$ = 0.20eV and $E_{ATOM}$ = 0.30eV. The values of $E_{CLUSTER}$ are given in the plot legend. 0.10ML coverage was deposited at a flux of 1.0MLs$^{-1}$ and the system was then allowed to relax for 0.10s.

We can see that when the island density decreases across the entire temperature range the greatest decrease occurs when the energy barrier for cluster attachment is greatest. This is because at low temperature individual mobile clusters are unlikely to attach to an immobile island and hence have a greater likelihood of being involved in a j cluster collision with other mobile clusters and nucleating a new island. As the temperature increases the clusters find it easier to attach to an immobile island and the likelihood of more than a single nucleation event occurring decreases along with the island density at equilibrium.

When $E_{CLUSTER}$ = 0.05eV and $E_{CLUSTER}$ = 0.10eV the island density initially increases with increasing temperature before decreasing at the highest temperatures. We can interpret this result by observing the morphology of the islands after the system has been allowed to relax, as shown in figure 15 for the case $E_{CLUSTER}$ = 0.05eV. We can see that at 500K several islands have nucleated and they have a fractal like shape. This is because when the energy barrier for cluster attachment is much lower than the energy barrier for adatom attachment the islands predominantly grow by the attachment of clusters. The islands' edges can only be smoothed out by the diffusion of single atoms and hence we are in a regime of effective irreversible aggregation by cluster attachment. Fractal islands are a familiar characteristic of irreversible aggregation in epitaxial systems.[19] At 550K there is only a single island present after relaxation due to the increased likelihood of clusters attaching to nucleated islands.

The interesting behaviour starts to occur at 600K. By observing the lattice we can see that some of the legs of the fractal island have broken off to become separate islands. This is due to the increased rate of detachment of atoms from islands and hence the island density at equilibrium increases. We can still see the remnants of the fractal nature of the larger island before break up occurred. At 650K the island density increases further due to the increased rate of break up of fractal islands but we can see that individual islands are better able to rearrange their shape through adatom diffusion. At 700K smaller islands are unable to survive the increased competition for adatoms due to increased rates of diffusion and hence the equilibrium island density begins to decrease again.

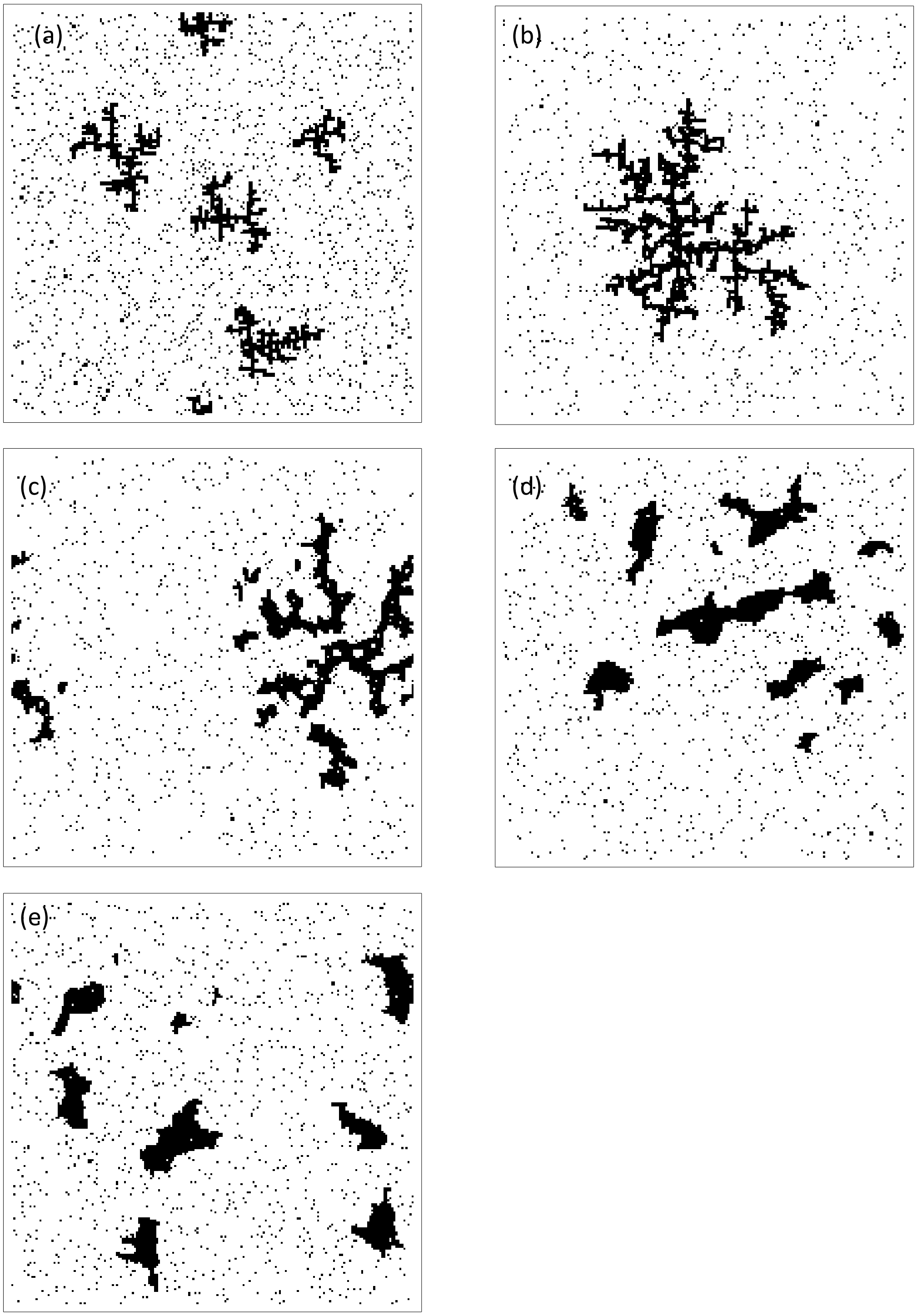

**Figure 15.** Images of lattices at (a) 500K, (b) 550K, (c) 600K, (d) 650K and (e) 700K with j=4. 0.10ML coverage deposited at 1.0MLs$^{-1}$ and then allowed to relax for a further 0.10s. Kinetic parameters are $E_D$ = 1.00eV, $E_{D'}$ = 0.80eV, $E_K$ = 1.40eV, $E_N$ = 0.20eV, $E_{ATOM}$ = 0.30eV and $E_{CLUSTER}$ = 0.05eV.

## 7.5 Growth velocity of islands

The original motivation for this work was provided by the non-linear edge velocity of graphene islands found during the LEEM experiments carried out by Loginova and co-workers.[14,15] The dependence of the edge velocity on the fifth power of the adatom supersaturation prompted the suggestion that graphene predominantly grows by the addition five atom clusters. It is therefore interesting to determine the edge velocity of islands formed in our KMC simulation of the tetramer model; in the case where we include a large energy barrier for adatom attachment to islands we would expect to find the step edge velocity depending on the fourth power of the adatom supersaturation.

Calculating the island edge velocity is simplest if we choose system parameters so that only a single island nucleation is likely and islands have an approximately circular shape. The equation for the island edge velocity is given by equation 2 and for the case of circular islands this reduces to,

$$v = \frac{dR}{dt} \qquad (15)$$

where R is the island radius. Figure 16 shows the island edge velocity as a function of adatom concentration for the tetramer model with j=6, $E_{CLUSTER}$ = 0.10eV and $E_{ATOM}$ = 0.30eV. A fit to the data of an edge velocity of the form suggested by Loginova and co-workers and given by equation 3 gives i=4.1±0.4. $n_{eq}$ = 0.032ML was determined from the simulation data and not used as a fitting parameter. This result is a neat check of the consistency of our tetramer model as it shows that the growth velocity of islands does indeed vary with the fourth power of the supersaturation of adatoms when the system parameters are such that we expect the growth of islands to be dominated by tetramer attachment.

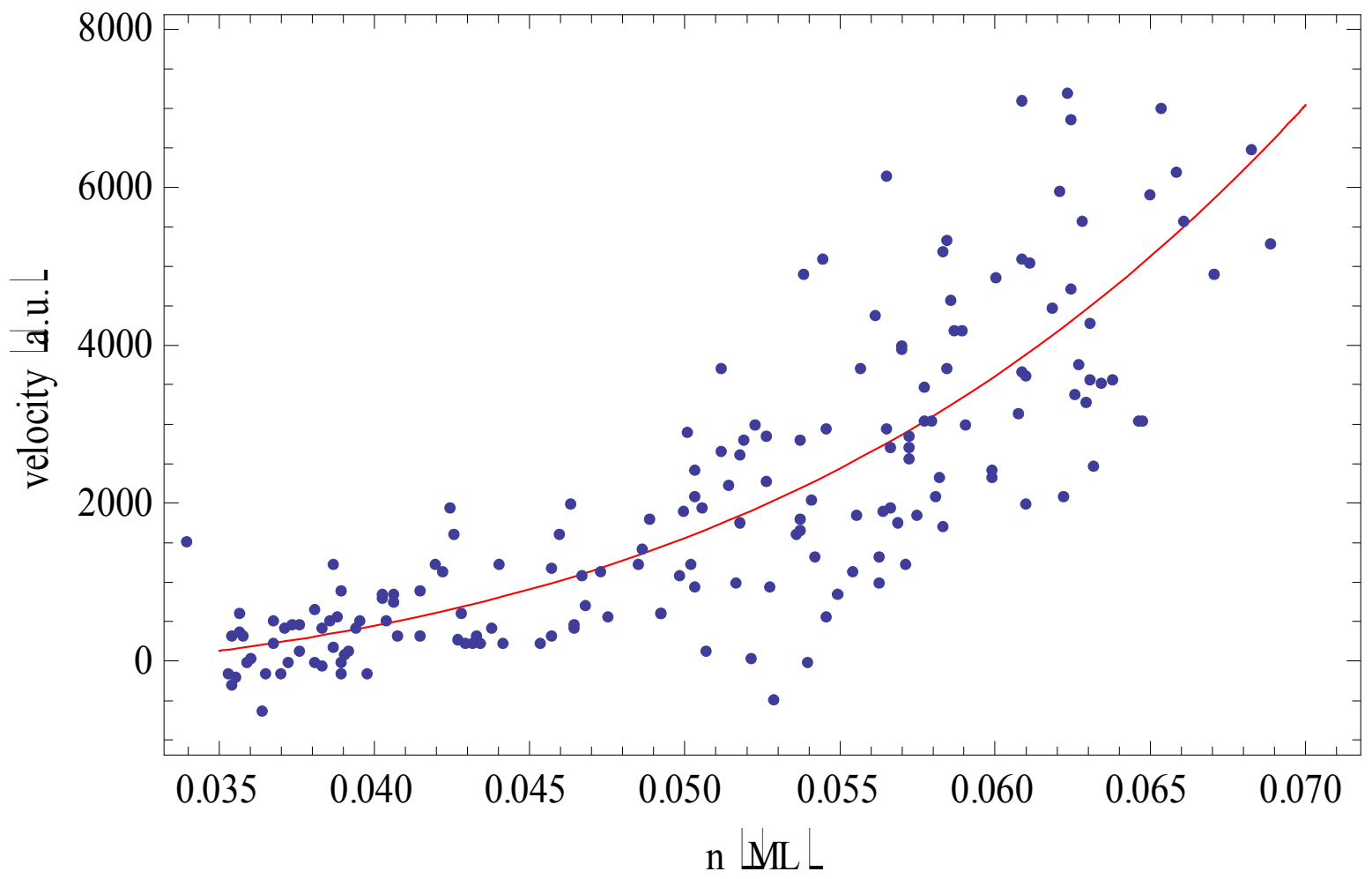

**Figure 16.** Island edge velocity (in arbitrary units) as a function of adatom density with j=6. The blue data points are data obtained from our simulations and the red curve is a fit to equation 3 with $n_{eq}$ = 0.032ML. Kinetic parameters are $E_D$ = 1.00eV, $E_{D'}$ = 0.80eV, $E_K$ = 1.40eV, $E_N$ = 0.20eV, $E_{ATOM}$ = 0.30eV and $E_{CLUSTER}$ = 0.10eV.

Figure 17 shows the island edge velocity as a function of adatom concentration for the tetramer model with j=6, $E_{CLUSTER}$ = 0.30eV and $E_{ATOM}$ = 0.30eV. A fit to the data of an edge velocity of the form given by equation 3 with $n_{eq}$ = 0.039ML gives i=4.4±0.8. For these parameters the island edge velocity is less well described by the equation suggested by Loginova and co-workers but the value found for i suggests that tetramer attachment is still the most significant process involved in island growth.

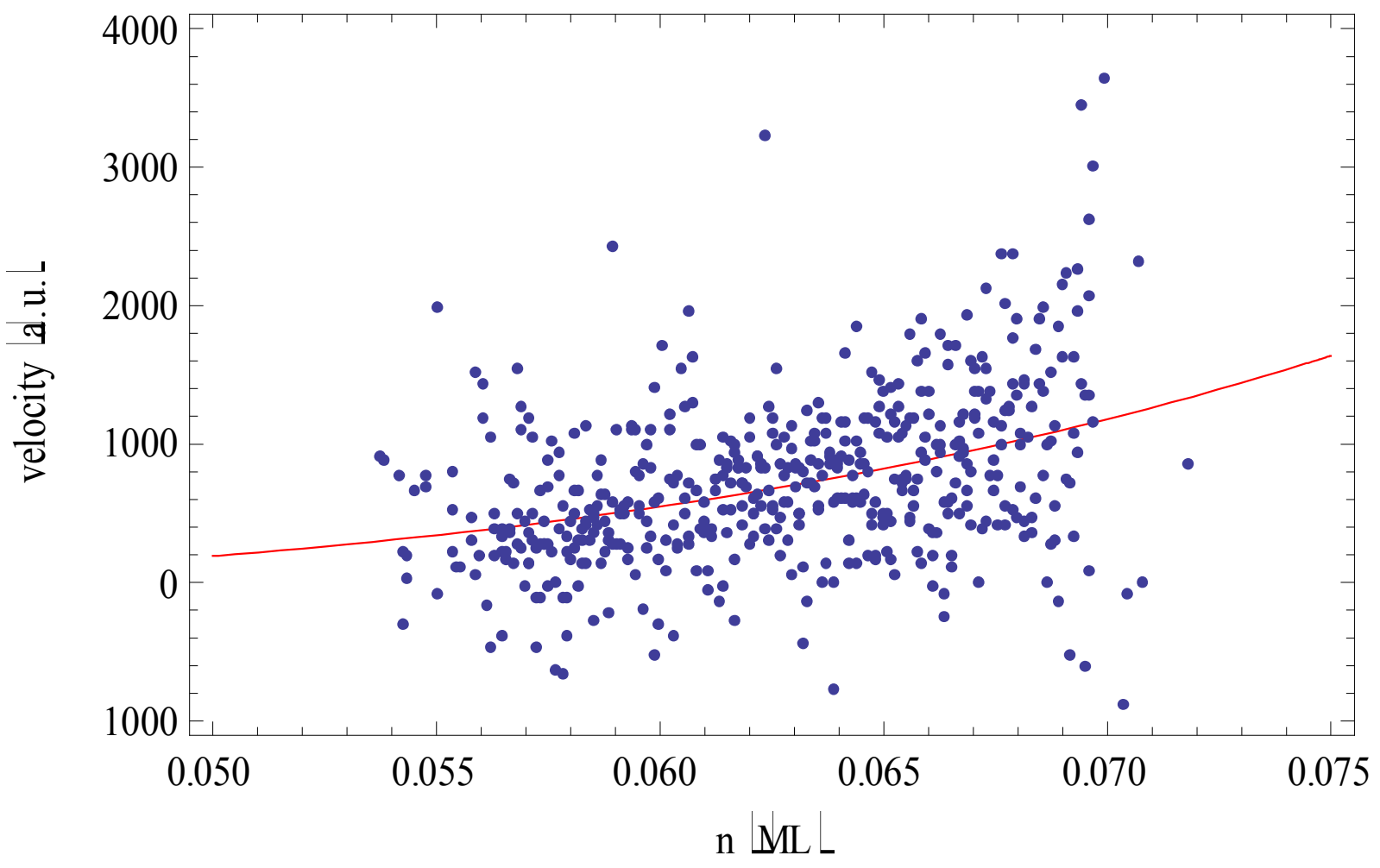

**Figure 17.** Island edge velocity (in arbitrary units) as a function of adatom density with j=6. The blue data points are data obtained from our simulations and the red curve is a fit to equation 3 with $n_{eq}$ = 0.039ML. Kinetic parameters are $E_D$ = 1.00eV, $E_{D'}$ = 0.80eV, $E_K$ = 1.40eV, $E_N$ = 0.20eV, $E_{ATOM}$ = 0.30eV and $E_{CLUSTER}$ = 0.30eV.

## 8. Errors

In order to obtain reliable data for the time evolution of the densities of adatoms, clusters and islands we took the mean values of the densities at each sampling point over a large number of runs using different seeds for the random number generator used in the simulations; the MT19937, which is available from the GNU Scientific Library.[28] The standard error, $\sigma_\mu$, of a sample mean estimate, μ, is given by,

$$\sigma_\mu = \frac{s}{\sqrt{n}} \qquad (16)$$

where s is the sample standard deviation and n is the size of the sample. In general we found that the fractional errors in the adatom and cluster densities are greatest around island nucleation and that the errors are larger at higher temperatures. We carried out enough simulation runs for each set of parameters so that the nucleation adatom density error was no greater than 5% and the nucleation cluster density error was no greater than 10%. The equilibrium island density error was no greater than 10%. Errors of this size were sufficiently small for our purpose as we were looking to observe qualitative trends in the data and not reproduce experimental results.

# 9. Conclusions

This report presents the results of Kinetic Monte Carlo simulations of a model of submonolayer epitaxial growth by the attachment of clusters consisting of multiple atoms. The initial motivation for this work was provided by the results of Low Energy Electron Microscopy experiments[14,15] carried out by Loginova and co-workers who investigated the nucleation and growth of graphene films on the transition metal ruthenium. The experiments showed that the growth velocity of 2D graphene islands varies with the fifth power of the supersaturation of carbon adatoms. This prompted the authors to suggest the graphene islands grow by the attachment of clusters of five atoms rather than by the usual mechanism of single adatom attachment. A microscopic interpretation of this result is that the energy barrier for single adatom attachment to graphene islands is prohibitively high and that in order for carbon atoms to attach to graphene an intermediate state consisting of multiple carbon atoms must form.

Zangwill and Vvedensky have developed a simple rate theory for the epitaxial growth of graphene on metal surfaces.[17] The novel feature of this model is that six mobile five atom clusters are required to collide in order to form graphene islands. The optimised set of rate equations which describe this model give a time dependent adatom density which is in quantitative agreement with experiment. The theory also well describes the temperature dependence of the nucleation and equilibrium densities of carbon adatoms. The nucleation density is an increasing function of temperature which is not typical of epitaxial systems and it was from this data that the result that six clusters are required to nucleate islands was obtained. The discrepancies that do exist between experiment and the rate theory are most likely due to the absence in the rate equations of spatial information which is important in epitaxial systems for determining the rate of adatom attachment and detachment from islands.

Kinetic Monte Carlo simulations explicitly take into account the spatial information and stochastic nature of epitaxial growth. We have carried out KMC simulations in order to further investigate the general scenario of epitaxial growth by the attachment of mobile clusters of atoms. In order to simplify our simulations we did not seek to directly replicate the graphene system but instead considered a model involving mobile tetramers on a square lattice. However we did wish to qualitatively account for three main features of the data from the LEEM experiments: the temperature dependence of the adatom density at the onset of island nucleation, the relationship between the values of the nucleation density and equilibrium density of adatoms and the temperature dependence of the equilibrium density of immobile islands.

The results of the simulations of the tetramer model show that the energy barrier for cluster break up and the number of tetramers required to nucleate an island are the important system parameters for determining whether the nucleation adatom density is an increasing function of temperature. $n_{nuc}$ increases across the entire range 550-700K when six tetramers are required to nucleate an island and with cluster break up energy barriers 1.20eV and 1.40eV. $n_{nuc}$ does not increase across the full temperature range for higher tetramer break up energies with j=6 and with any tetramer break up energies for j=2 and j=4. When six tetramers are required to nucleate an island the effect of the increasing rate of tetramer diffusion with increasing temperature is completely negated by the increasing rate of tetramer break up and hence the density of adatoms required to nucleate an island increases with temperature.

The results of the simulations of our original tetramer model show that the nucleation density of adatoms is an order of magnitude greater than the adatom density at equilibrium with islands and clusters. For the graphene system Loginova and co-workers reported that $n_{nuc} \approx 2n_{eq}$.[14,15] When we added an additional energy barrier for adatom attachment to islands to our model then $n_{eq}$ increases without affecting the value of $n_{nuc}$. A relatively high energy barrier of at least 0.30eV is required in our model to give $n_{nuc} \approx 2n_{eq}$ across the range 550-700K.

Loginova and co-workers[14,15] reported that the graphene island density dramatically decreases with increasing temperature on Ru(0001). In order for our tetramer model to display this behaviour we need to add an energy barrier for tetramer attachment to islands. For an adatom attachment barrier of 0.30eV a tetramer attachment barrier of greater than 0.15eV is required for the equilibrium island density to decrease across the range 500-700K. The greatest decrease in island density is shown for tetramer attachment energy 0.25eV because at the lowest temperature tetramer find it difficult to attach to existing islands and have a greater likelihood of being involved in a j tetramer collision and nucleating a new island. For tetramer attachment energy 0.05eV and 0.10eV the equilibrium island density initially increases with increasing temperature. This is because of the break up of fractal islands into a number of smaller islands by adatom detachment.

A fit to data from our simulations of an island edge velocity of the form suggested by Loginova and co-workers[14,15] shows that, when tetramer attachment is the dominant process for island growth, islands grow with a velocity that varies with the fourth power of the supersaturation of adatoms.

We now consider what further work could be carried out to advance our understanding of epitaxial graphene growth.

The sub-monolayer island size distribution of certain epitaxial systems has been investigated experimentally, for example for Fe/Fe(001) homoepitaxy,[29] and KMC simulations have been carried out to find energy parameters that reproduce the experimental data using the experimental temperature and flux values.[24] The LEEM experiments carried out by Loginova and co-workers facilitate the calculation of island size distribution for the Gr/Ru(0001) system. Although we would not be able to replicate the LEEM experimental conditions, our tetramer model could be used to try and qualitatively reproduce the main features of the measured graphene island size distribution. Some of our most recent results showing the island size distribution for the tetramer model are included in the appendix of this report.

Measurements of the island size distribution during equilibration once the adatom flux is turned off could also reveal if $n_{eq}$ is predominantly determined by adatom detachment and then re-attachment to the same island. This would help to confirm the suggestion made by Zangwill and Vvedensky as to why the adatom density obtained from the rate equations agrees best with experiment at equilibrium.

In this work we have not attempted to replicate with quantitative accuracy any experimental results of epitaxial graphene growth. The next step in seeking to understand graphene growth would be to carry out KMC simulations of a model of epitaxial growth by five atom cluster attachment on a hexagonal lattice. If sufficient computing power became available then the rate equation results of Zangwill and Vvedensky suggest that this model could replicate the results of the LEEM experiments carried out by the Sandia group.

An alternative approach to model graphene growth under realistic experimental conditions is the subject of a proposed PhD project at the Theory and Simulation of Materials Centre for Doctoral Training (TSM CDT).[30] This will involve developing lattice-free kinetic Monte Carlo simulations of graphene growth on metal surfaces with simulation parameters obtained from first principles density functional theory calculations.

So far lattice-free KMC simulations have only been carried out in (1+1) dimensions.[31,32] The aim of such simulations is to overcome the limitations of a fixed lattice and introduce a pair potential $U_{ij}$ between two atoms i and j which are separated by a continuous distance $r_{ij}$. The Lennard-Jones potential has been used successfully to gain insight into mechanism of heteroepitaxial growth.[32] In lattice-free KMC the activation energy for a diffusion step is given by the difference between the energy of an atom in the transition state and its energy in the binding state. In (1+1) dimensions the transition state / binding state energies are found by maximising / minimising the energy of an atom with respect to its vertical coordinate. The PhD project at the TSM CDT will develop a 2D version of this method.

## 10. Acknowledgements

I would like to thank my supervisor, Professor Dimitri Vvedensky, and my project partner, Bartomeu Monserrat, for the many valuable and informative discussions that have taken place during the course of this project.

As mentioned in section 4, both Bartomeu Monserrat and I have written independent versions of the code that simulates the tetramer model and hence we are both capable of producing all the data reported here. However, due to time constraints, we split up the running of the code that produced our final results. I would therefore like to thank Bartomeu Monserrat for providing the data displayed in figures 11b, 11c, 12b, 12c, 13, 16 and 17.

# Appendix

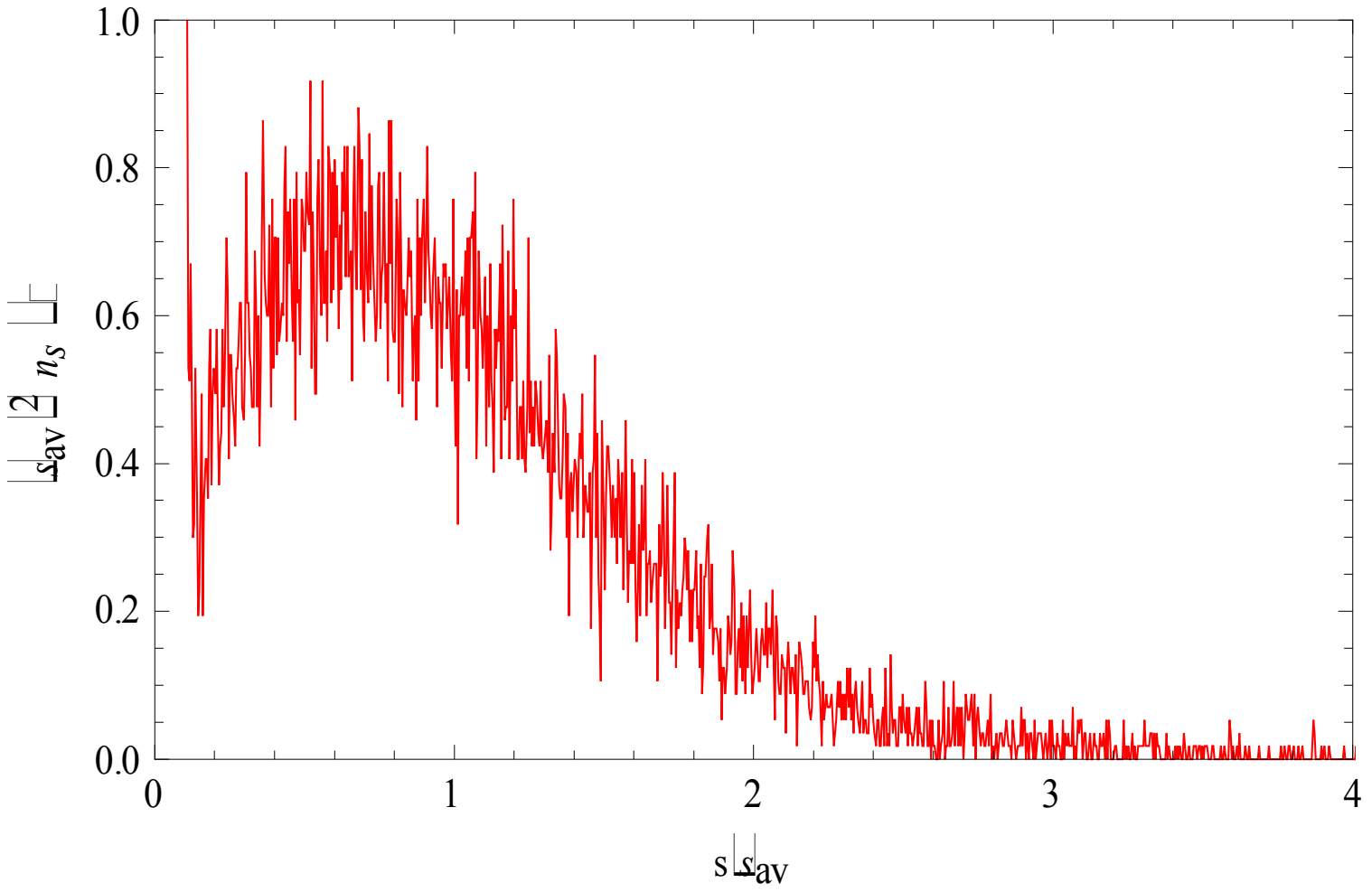

**Figure A1.** Scaled island size distribution at 500K for j=6. Kinetic parameters are $E_D$ = 1.00eV, $E_{D'}$ = 0.80eV, $E_K$ = 1.40eV, $E_N$ = 0.20eV, $E_{ATOM}$ = 0.30eV and $E_{CLUSTER}$ = 0.25eV. 0.20ML coverage was deposited at a flux of 1.0MLs$^{-1}$ and the system was then allowed to relax until equilibrium was reached. Data provided by Bartomeu Monserrat.